\documentclass[12pt]{article}
\pdfoutput=1

\usepackage{paper2e}
\usepackage{mydefs2e}

\usepackage{xspace}
\usepackage{graphicx}

\newcommand{\gap}{\hspace{0.05em}}

\newcommand{\mq}{m_{Q}}
\newcommand{\muq}{\mu_{Q}}

\newcommand{\LQ}{\La_{\rm QCD}}
\renewcommand{\d}{\partial}

\begin{document}

\begin{titlepage}

\title{Macroscopic Strings and ``Quirks''\medskip\\
at Colliders}

\author{Junhai Kang,$^{\rm a}$\ \ Markus A. Luty,$^{\rm a,b}$}

\address{$^{\rm a}$Physics Department, University of Maryland\\
College Park, Maryland 20742}

\address{$^{\rm b}$Physics Department, University of California Davis\\
Davis, California 95616}

\begin{abstract}
We consider extensions of the standard model containing
additional heavy particles (``quirks'') charged under a new 
unbroken non-abelian gauge group as well as the standard model.
We assume that the quirk mass $m$ is in the phenomenologically
interesting range $100\GeV$--TeV,
and that the new gauge group gets strong at a scale $\La < m$.
In this case breaking of strings is exponentially suppressed,
and quirk production results in strings that are long compared to
$\La^{-1}$.
The existence of these long stable strings leads to highly
exotic events at colliders.
For $100\eV \lsim \La \lsim \mbox{keV}$ the strings
are macroscopic, giving rise to events with two
separated quirk tracks with measurable curvature
toward each other due to the string interaction.
For $\mbox{keV} \lsim \La \lsim \mbox{MeV}$ the typical
strings are mesoscopic: too small to resolve in the detector,
but large compared to atomic scales.
In this case, the bound state appears as a single particle,
but its mass is the invariant mass of a quirk pair, which
has an event-by-event distribution.
For $\mbox{MeV} \lsim \La \lsim m$, the strings
are microscopic, and the quirks annihilate promptly
within the detector.
For colored quirks, this can lead to hadronic fireball
events with $\sim 10^{3}$
hadrons with energy of order GeV
emitted in conjunction with hard decay products from the final
annihilation.
\end{abstract}

\end{titlepage}

\section{Introduction}
\label{sec:intro}
The LHC has energized the particle physics community 
with the promise of new physics at the TeV scale.
This is the scale where the origin of electroweak symmetry
breaking and the solution of the hierarchy problem must lie.
Most studies of physics beyond the standard model therefore involve
minimal models directly motivated by these problems,
most notably the MSSM.
However, history teaches us that the true physics may be non-minimal,
and the most striking experimental discoveries may not have any
obvious connection to the ``big questions.''
It is therefore important to look for any new physics that can
manifest itself by the enhanced energy reach of a new accelerator,
especially if it arises from a simple extension of the standard model.
This is especially important at a hadron collider such as the LHC,
where large backgrounds mean that finding a signal often requires
knowing what to look for.

The classic example of simple new physics not directly motivated
by electroweak symmetry breaking is a $Z'$.
This involves extending the standard model with a $U(1)'$ gauge
group, plus a new Higgs sector that breaks the $U(1)'$ symmetry.
The only parameter of the new Higgs sector relevant for
phenomenology is the $Z'$ mass, so the only parameters in the model
are the $U(1)'$ coupling $g'$ and $m_{Z'}$.
In addition, there is a discrete choice of the charges of standard
model fields under $U(1)'$.
(We assume that some of these charges are nonzero,
otherwise the quirks are not observable.)
Although the $Z'$ mass is not directly tied to electroweak
symmetry breaking, the focus is on the phenomenologically interesting
regime (very roughly $m_{Z'} \sim \mbox{TeV}$) that is not excluded by
existing experiments, but may be probed at LHC.

In this paper, we consider another equally minimal
extension of the standard model.
We assume that there is an additional unbroken $SU(N)$
gauge group
with some fermions $Q$, $\bar{Q}$ in the fundamental representation.
(The qualitative features of the model are unchanged 
if the particles are scalars rather than fermions.)
This model is parameterized by the mass of the new particles
$\mq$ and the $SU(N)$ gauge coupling, which can be parameterized
by the scale $\La$ where it gets strong.
In addition, there is a discrete choice of the standard model
gauge quantum numbers of the new fermions.
(We assume that some of these charges are nonzero,
otherwise the $Z'$ is not observable.)
We assume that the mass of the fermions is in the phenomenologically
interesting range (very roughly $100\GeV \lsim \mq \lsim \mbox{TeV}$)
that is not excluded by
existing experiments, but may be probed at LHC.

New strong interactions have been considered often in particle physics,
usually with strong interaction scales at or above a TeV.
We instead consider the case where $\La \ll \mbox{TeV}$, in particular
\beq
\La \ll \mq.
\eeq
We therefore refer to the new gauge interaction as ``infracolor.''
Note that if $Q$ is the lightest particle in the fundamental
representation of infracolor then it is automatically stable,
since there is no lighter state with the same quantum numbers.
We have learned recently that this model was first considered
by L.\ B.\ Okun \cite{thetons}, who called the new particles
``thetons.''
This model was considered as a limit
of QCD in \Ref{QCDlongstrings}.
\Ref{HV} also mentioned this model as an example of
a ``hidden valley'' model.

This paper will consider the phenomenology of these models
with values of $\La$ ranging over many orders of
magnitude (roughly $100\eV$ to $100\GeV$).
All these values are natural,
since $\La$ is related to the fundamental
gauge coupling $g_0$ defined at a scale $\mu_0$ by
\beq
\La = \mu_0 \, e^{-8\pi^2 / b g_0^2},
\eeq
where $b$ is the 1-loop coefficient of the $SU(N)$ gauge coupling
beta function.
The scale $\La$ is exponentially sensitive to the value of
$g_0^2$, so each decade of energy is roughly equally likely.

Cosmology places strong constraints on new light physics,
even if it is weakly coupled to the standard model.
However, if the reheat temperature is sufficiently low
($T \lsim \mbox{GeV}$) the infracolor sector is never in thermal
equilibrium, and there are no cosmological consequences.
This shows that
there are no model-independent constraints from cosmology
on this physics.
If we assume thermal abundances for the new particles
the cosmology is complicated, but may also be viable
\cite{quirkcosmo}.

This paper will focus on the collider phenomenology of this model
at the qualitative level.
This phenomenology of this simple model is surprisingly exotic.
The reason is that breaking of the infracolor gauge string is
exponentially suppressed due to the large $Q$ mass.
As we will see, this leads to very exotic phenomenology,
so we call the new particles ``quirks.''%
\footnote{%
This can also be motivated by the replacements
$\mbox{``strong''} \to \mbox{``string''}$,
$\mbox{``quark''} \to \mbox{``quirk''}$.
}

The collider phenomenology of quirks depends crucially on the length
of the strings.
This is set by the scale where the quirk kinetic energy
is converted to string potential energy.
Since the typical event is not close to threshold, it has kinetic
energy $\sim \mq$ and gives a string length scale
\beq
L \sim \frac{\mq}{\La^2} \sim 10~\mbox{m}
\left( \frac{\mq}{\mbox{TeV}} \right) 
\left( \frac{\La}{100\eV} \right)^{-2}.
\eeq
We will consider collider signals for string
length scales ranging from 
the size of detectors ($\sim 10~\mbox{m}$)
to microscopic scales.

This paper is organized as follows.
In Section \ref{sec:models}, we briefly
discuss model-building issues
such as naturalness and unification, as well as indirect
constraints from precision electroweak data, cosmology,
and astrophysics.
In Section 3
we discuss production of quirks and strings.
In Section 4
we discuss signals for macroscopic strings.
In Section 5
we consider annihilation of quirks catalyzed by
the string.
In Section 6
we discuss the signals of mesoscopic strings,
those that are too small to be resolved in a detector
but large compared to atomic scales.
In Section 7
we discuss the collider signals from microscopic strings.
Section 8 contains our conclusions.

\section{Models and Indirect Constraints}
\label{sec:models}
In this section, we discuss model-building issues such as
naturalness and unification, as well as
indirect constraints from precision electroweak constraints
and cosmology.
This discussion is fairly standard, and our conclusion is that
there are no strong model-independent constraints on quirks from
these considerations.

\subsection{Coupling to the Infracolor Sector}
Because we assume that the scale of infracolor strong interactions
is below the weak scale, the hadrons of the infracolor sector are
kinematically accessible to existing experiments.
However, the standard model is uncharged under infracolor,
and therefore a quirk loop is required to couple the sectors.
Since the quirks are heavy, this leads to highly suppressed couplings
to the infracolor sector.

\begin{figure}[t!]
\begin{center}
\includegraphics[]{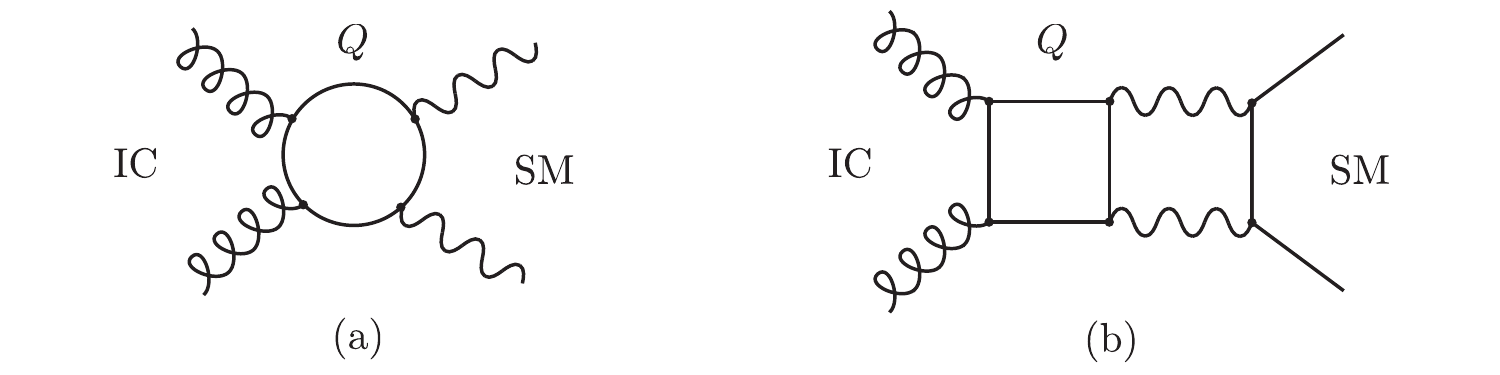}
\begin{minipage}[t]{5in}%
\caption[]{Loop graphs contributing to the coupling of the
standard model and infracolor sector.}
\label{fig:SMICloop}
\end{minipage}
\end{center}
\end{figure}

The leading coupling between the standard model and the 
infracolor sector at low energies
arises from the diagram of Fig.~\ref{fig:SMICloop}a.
This gives rise to the dimension-8
effective operator
\beq[LeffICSM]
\scr{L}_{\rm eff} \sim \frac{g^2 g'^2}{16\pi^2 \mq^4}
F_{\mu\nu}^2 F^{\prime 2}_{\rho\si}.
\eeq
The 2-loop diagram of Fig~\ref{fig:SMICloop}b can couple the infracolor
gauge fields to dimension-3 fermion bilinears, but these
have an additional helicity suppression in addition to
the additional loop suppression, and are therefore suppressed.
For $\mq \gsim 100\GeV$ this operator is far weaker than the
weak interactions, so production of infracolor gauge bosons
at colliders
with energy below the quirk mass is completely negligible.
Probing this sector at colliders requires sufficient energy
to produce quirks directly.

The operator \Eq{LeffICSM} mediates infracolor glueball decay,
for example to photons or gluons.
The rate is of order
\beq
\Ga \sim \frac{1}{8\pi} \left( \frac{g^2 g'^2}{16\pi^2 \mq^4} \right)^2
\La^9.
\eeq
Note that this is very sensitive to both $\La$ and $\mq$.
We have
\beq[icglueballdecay]
c\tau \sim 10~\mbox{m}\ \left( \frac{\La}{50\GeV} \right)^{-9}
\left( \frac{\mq}{\mbox{TeV}} \right)^{-8}.
\eeq
We see that the infracolor glueballs can decay inside a particle
detector for $\La \gsim 50\GeV$, while the lifetime becomes longer
than the age of the universe for $\La \lsim 50\MeV$.

\subsection{Star Cooling}
Stars with temperature $T \gsim \La$ can potentially cool due
to emission of infracolor glueballs.
Due to the rapid decoupling of infracolor interactions from
standard model interactions in \Eq{LeffICSM},
we find that this does not give interesting bounds.

We will focus on bounds from SN1987A,
which has the highest temperature ($T \sim 30\MeV$)
of the astrophysical systems used to constrain light particles.
We can estimate the bounds by comparing to axion cooling,
which constrains the axion decay constant
$f_a \gsim 10^9\GeV$.
For both the axion and infracolor, the dominant energy loss mechanism
is nuclear bremmstrahlung.

Below the QCD scale the coupling \Eq{LeffICSM}
gives rise to an effective coupling of infracolor
gauge fields to nucleons:
\beq
\scr{L}_{\rm eff} \sim \frac{g^2 g'^2 \LQ}{16\pi^2 \mq^4}
\bar{N} N F'^2_{\mu\nu}.
\eeq
Here $\LQ \sim 1\GeV$ is the scale of strong QCD interactions.
Factors of $4\pi$ have been put in using ``\naive\ dimensional
analysis'' \cite{NDA}.
This is to be compared with the axion coupling
\beq
\scr{L}_{\rm eff} \sim \frac{\LQ}{f_a} a \bar{N} N.
\eeq
We therefore have
\beq
\frac{\rm rate\ of\ infracolor\ production}{\rm rate\ of\ axion\ production}
\sim \left. 
\left(\frac{g^2 g'^2 \LQ}{16\pi^2 \mq^4}\right)^2 T^6
\right/ 
\frac{\LQ^2}{f_a^2},
\eeq
which gives a bound of
$\mq \gsim 10\GeV$.
Although these estimates are very crude, the fact that
the infracolor emission falls as $1/\mq^8$ means that
the rate is highly suppressed in the interesting regime
$\mq \gsim 100\GeV$.

\subsection{Cosmology}
If the infracolor gauge interactions and/or the quirks have
thermal abundances early in the universe, 
there are stringent cosmological constraints.
This paper will focus mainly on collider physics, so we make
only some simple remarks here, leaving a more complete
investigation to future work.

The rapid decoupling of the infracolor interactions means that
infracolor glueballs are not produced if the reheating
temperature is sufficiently low.
Assuming $T_{\rm RH} \gg \La$, the condition for infracolor
interactions to be out of equilibrium is
\beq
\Ga \sim \left( \frac{g^2 g'^2}{16\pi^2 \mq^4} \right)^2 T_{\rm RH}^9
\gsim \frac{T_{\rm RH}^2}{M_{\rm P}},
\eeq
which is satisfied for $T_{\rm RH} \lsim \mbox{GeV}$.
This is easily sufficient for nucleosynthesis at $T \sim \mbox{MeV}$,
the highest temperature about which we have secure cosmological
knowledge.

This is not an entirely satisfactory solution to cosmology,
since it requires dark matter and the baryon asymmetry to be produced
at low termperatures.
This is possible with \eg\ MeV dark matter \cite{MeVDM}
and low-scale baryogensis \cite{lowscaleBG}.
We can avoid exotic low-temperature cosmology
by having quirks decay
to infracolored states that are sterile under the standard
model.
These decays can have lifetimes long compared to collider
time scales, but short enough to avoid cosmological constraints.
We will not discuss the details here.
For the present discussion it is sufficient
that low reheat temperatures are not in conflict with
nucleosynthesis, so there is no model-independent
constraint from cosmology.

\subsection{Precision Electroweak Data}
Precision electroweak data constrains new physics
at the TeV scale.
However, if the quirks are in a vector-like representation of the standard
model gauge group they can have a TeV mass term that does not
break electroweak symmetry.
Furthermore, virtual quirks are necessarily created in pairs, 
so there are no tree-level effects on electroweak observables.
There is therefore no constraint on such models from precision
electroweak data.

\subsection{Model Building}
Next we discuss the plausibility of this kind of new physics.
The existence of additional gauge groups with matter in
bifundamental representations is a hallmark of brane constructions
in string theory.
As we will see the most natural quirk sector is vectorlike,
which means that it requires no additional projections of the
kind needed to obtain a chiral theory such as the standard model. 
A quirk/infracolor sector can therefore arise simply and
naturally from string theory.

In fact, in realistic supersymmetric theories there is already at least
one set of vectorlike fields, namely the Higgs bosons.
These must have a supersymmetric ``$\mu$ term'' at the weak scale,
otherwise we have either light Higgsinos or no electroweak symmetry
breaking.
Any mechanism that generates the $\mu$ term can also generate a weak-scale
mass for the quirks.
This means that no additional assumptions are required to explain
the origin of the quirk mass in supersymmetric theories.
It is also trivial to preserve gauge coupling unification in
supersymmetric theories by assuming that the quirks come
in complete GUT representations.
The simplest example is that the quirks are in a
\beq[fiveplusfivebar]
{\bf 5} \oplus \bar{\bf 5}
\to ({\bf 3}, {\bf 1})_{\frac 13}
\oplus (\bar{\bf 3}, {\bf 1})_{-\frac 13}
\oplus ({\bf 1}, {\bf 2})_{\frac 12}
\oplus ({\bf 1}, {\bf 2})_{-\frac 12}.
\eeq
In this model there is no
tree-level Yukawa interaction that can split the masses
of the doublet.
These splittings will arise from loop graphs, and will be very small.
There is also no tree-level interaction that allows either
the color triplet or the electroweak doublet to 
decay to the other, so this model naturally has both
colored and uncolored quirks.

In fact, a quirk/infracolor sectors have already
appeared in some model-building
motivated by the hierarchy problem.
Such a sector was proposed in \Ref{BKG}
to give additional loop contributions to the physical Higgs mass
in supersymmetry.
Scalar quirks (``squirks'') appear in models of ``folded supersymmetry''
\cite{foldedSUSY}.

Small values of $\La$ are perfectly
compatible with grand unification.
As an example, we consider the MSSM with an $SU(2)$
infracolor gauge group, with quirks in the
${\bf 5} \oplus \bar{\bf 5}$ representation (see \Eq{fiveplusfivebar}).
The infracolor beta function is equal to the color beta function
at one loop, simple unification implies that
the infracolor gauge coupling is equal to the 
QCD gauge coupling at the scale of superpartner masses.
The scale of infracolor interactions is then of order $100\MeV$.
If the theory above the TeV scale has
respectively 1, 2, 3 additional pairs of infracolor fundamentals,
the infracolor scale is respectively MeV, $10\keV$, $100\eV$.

\section{Quirk Production and String Formation}
\label{sec:production}

\subsection{Absence of String Breaking}
\begin{figure}[t!]
\begin{center}
\includegraphics[]{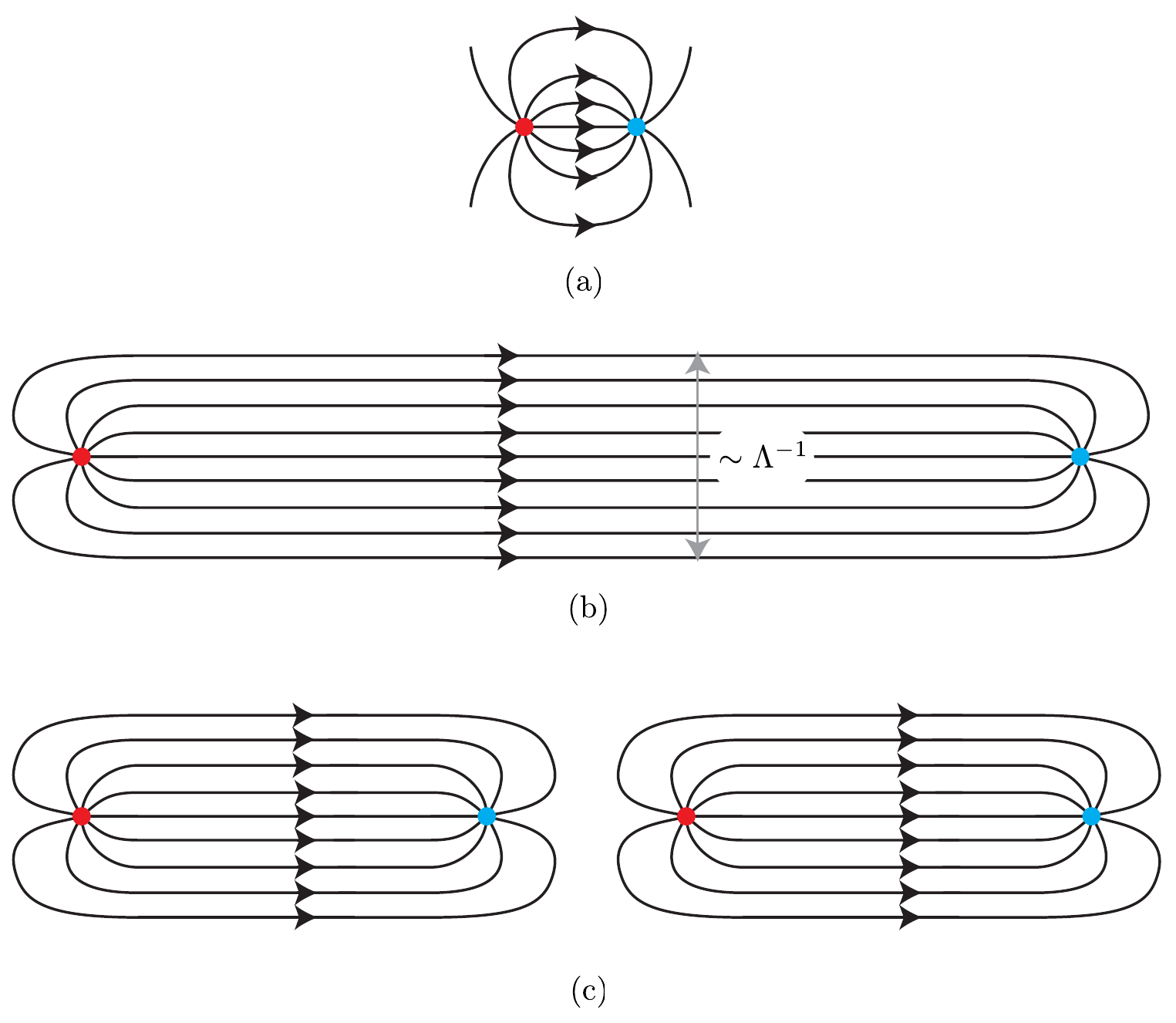}
\begin{minipage}[t]{5in}%
\caption[]{Schematic view of color flux for
quirk separation for (a) $r \ll \La^{-1}$ and
(b) $r \gg \La^{-1}$.
String breaking (c) requires a quirk-antiquirk pair to be
created, which costs energy $2\mq \gg \La$.}
\label{fig:stringbreak}
\end{minipage}
\end{center}
\end{figure}

The reason that infracolor gauge strings do not break was already
discussed in the introduction.
A virtual quirk-antiquirk pair costs energy at least $2\mq$,
and will have a typical separation of order $\mq^{-1}$.
This lowers the string potential energy only by an amount
of order $\La^2 / \mq \ll 2\mq$, so this process cannot
go on shell (see Fig.~\ref{fig:stringbreak}).
An on-shell transition requires eliminating a length
of string of order $\De L \sim \mq / \La^2 \gg \La^{-1}$.
The rate for this transition will be exponentially suppressed.

This transition is closely analogous to the Schwinger
mechanism of pair creation of charged particles by a
weak electric field \cite{stringbreak}.
For charged particles with
$m \gg E^{1/2}$, the rate for pair creation is
\beq
\Ga/V = \frac{E^2}{4\pi^3} e^{-\pi m^2 / E}.
\eeq
Modeling a gauge string as a flux tube with area $A$
and color electric field $E$, the string tension is
$\si \sim E^2 A$, so we have
\beq
\Ga/L \sim \frac{\si}{4\pi^3} e^{-\pi \mq^2 / E}.
\eeq
For a string of length $L \sim \mq / \si$,
we estimate $E \sim \pi \La^2$ and obtain a lifetime
\beq
\tau \sim \frac{4\pi^3}{m_Q} e^{m_Q^2 / \La^2}.
\eeq
For $\La/\mq \lsim 10^{-1}$ this is already
longer than the age of the universe for $\mq \gsim 100\GeV$.

\subsection{Quirk Production}
\begin{figure}[t!]
\begin{center}
\includegraphics[]{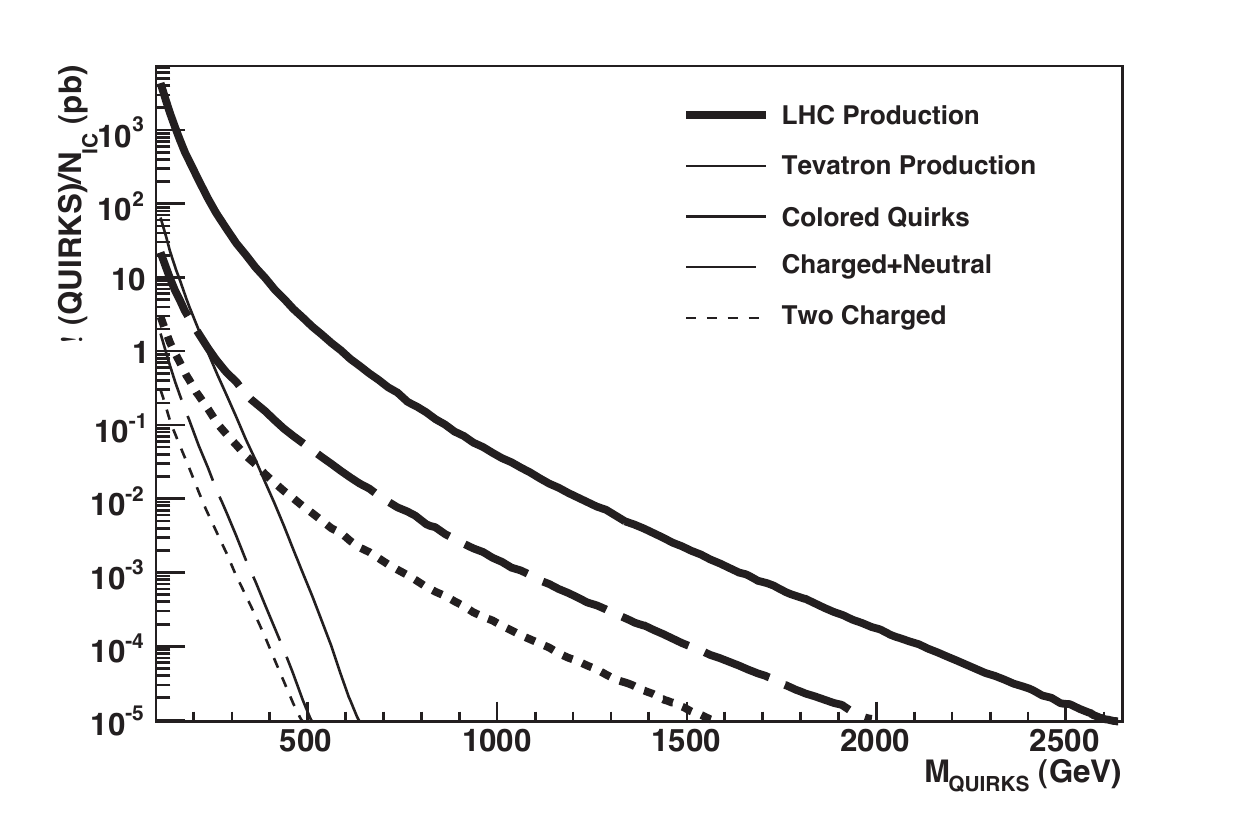}
\begin{minipage}[t]{5in}%
\caption[]{Quirk production cross section
at the Tevatron and LHC.}
\label{fig:Qprod}
\end{minipage}
\end{center}
\end{figure}

Quirk production involves energy and momentum transfer
of order $\mq \gg \La$ and $\LQ$, and is therefore a hard
perturbative process.
The total cross section for quirk production at Tevatron and LHC
at leading order in perturbation theory are shown in
Fig.~\ref{fig:Qprod}.
This does not include Sommerfeld enhancement due to attractive
infracolor and/or QCD interactions
\cite{Sommerfeld}.
This will increase the cross section near threshold,
and need to be included in a more detailed study.
The Coulomb interactions are familiar, so we consider briefly the 
Sommerfeld enhancement due to the long-range infracolor interactions.
These become relevant only when the string length is longer than
$\La^{-1}$, which requires $\be \gsim (\La / \mq)^{1/2}$.
The linear potential will be a large perturbation on the state
if the potential energy changes significantly in one de Broglie
wavelength.
We therefore compute the ratio of this change to the kinetic energy:
\beq
\frac{\De V}{K} \sim \frac{\La^2 / \mq\be}{\mq \be^2}
\sim \frac{\La^2}{\mq^2}\, \frac{1}{\be^3}
\lsim \left( \frac{\La}{\mq} \right)^{1/2}.
\eeq
We see that the effects of the long-range potential are always
small enough to be treated as a perturbation,
although they may be numerically significant
for the largest values of $\La$.

Returning to Fig.~\ref{fig:Qprod},
we conclude that the cross sections are substantial up to several
TeV for LHC. 
(Note that the cross section is proportional to $N_{\rm IC}$.)
Many quirk signatures are completely background-free 
(as we will see), so even a few reconstructed events may be
sufficient for discovery.

\subsection{String Formation}
\label{subsec:stringprod}
The effect of non-perturbative infracolor interactions and the
formation of an infracolor string has many points in common
with hadronization of heavy stable quarks in QCD,
so we review this first.

Imagine that there is a heavy ($m \gg \LQ$) stable quark
(or squark or gluino) in QCD in addition to the light quarks.
Below the free quark threshold at $2m$ these can be produced in
a Coulomb-like bound state (quarkonium).
Formation of such a low-lying bound state requires that the
quirk pair be produced just below the free threshold, \ie
$2m - E \sim \al_3^2(m) m$.
Wavefunction overlap factors give a
suppression of the rate by additional
powers of $\al_3(m)$, and so the rate for the production of
these bound states is much smaller than the production
rate for unbound quarks.
The quark production cross section is dominated by 
quarks with kinetic energy $K = E - 2m \sim m \gg \La$.
In this regime, threshold effects are unimportant
and the production process is perturbatively
calculable in an expansion in $\al_3(m)$. 

We now consider the hadronization of heavy stable quarks
with $K \sim m$.
Kinematically, it is possible that a large fraction of the kinetic
energy is converted to light hadrons, resulting in a jet
surrounding the heavy quarks.
However, because the quark is very heavy its kinetic energy cannot
be efficiently converted into production of light hadrons.
The basic reason is that the strong interactions have a
range of order $\LQ^{-1}$, so once the heavy quarks are separated
by a distance $r \gg \LQ^{-1}$ the strong interactions become
perturbative.
It is traditional in heavy quark physics to refer to the
non-perturbative QCD interactions as ``brown muck'' to
emphasize how little we know about it.
The size of the force exerted by the brown muck is of
order $\LQ^2$, so the total energy transfered from quark kinetic
energy into light hadrons is only of order
\beq
\De E \sim F \De r \sim \LQ.
\eeq
There is a tail at large $\De E$ that can be described
in perturbative QCD by additional hard gluons.

We now turn to quirks.
The infracolor interactions
effectively have infinite range because of the infracolor string,
and we might worry that the conversion of quirk kinetic
energy to infracolor hadrons (glueballs) never stops.
We consider events far from threshold ($K \sim \mq$), for
which the string length $L \sim \mq / \La^2 \gg \La^{-1}$,
long enough to be a well-defined object.
In this case the string rapidly straightens out,
approaching a configuration close to its local ground state.

To understand this, it is helpful to restate in a somewhat
formal way the obvious fact that well-separated QCD hadrons
from heavy stable quark production do not continue to lose energy
to hadron emission.
The point is that a state consisting of well-separated hadrons
is locally (on scales of order $\LQ$)
a boost of the ground state.
Let us apply this point of view to a rapidly-stretching infracolor string
with heavy quirks at the ends.
In the center of mass frame, the middle of the string
has zero transverse velocity.
A long QCD string is described by the Nambu-Goto action (see below),
which has no longitudinal excitations.
This string configuration is therefore identical to the 
ground state in the center of mass frame.
Near the ends of the string, only the acceleration of the ends
represents a departure from a boost of a ground state.
The acceleration is given by
\beq
a = \frac{F_{\rm string}}{\mq} \sim \frac{\La^2}{\mq} \ll \La.
\eeq
Because the acceleration is very small on the scale $\La$,
there is no energy loss to infracolor radiation from the ends.
The infracolor strings can be thought of as being close to the static
limit $\mq \to \infty$.
This is qualitatively different from the open strings of string
theory, which have massless ends.

The non-perturbative infracolor ``brown muck'' is therefore
effective in radiating glueballs only when the quirk separation is
of order $\La^{-1}$ or less.
Similarly to the case of heavy stable quark production in QCD,
this results in an energy of order $\La$ being radiated
into infracolor glueballs during the production process.

\subsection{Dynamics of Quirks and Strings}
We now discuss the motion of the quirk-string system
produced as described above.
As long as we are considering excitations of the string
with wavelengths long compared to $\La^{-1}$, we can use
an effective description in which the string is elementary.
This is analogous to the chiral Lagrangian describing pion
interactions for energies small compared to $\LQ \sim \mbox{GeV}$.

Gauge strings are described at long distances
by the Nambu-Goto action.
This is not \emph{a priori} obvious, since there are other
universality classes of strings that break additional Lorentz
symmetry.
For a clear discussion of this point, see \Ref{Sundrumstring}.
Strong numerical evidence that the long-wavelength
fluctuations of the QCD string are
described by the Nambu-Goto action was obtained in \Ref{Luscher}.

\begin{figure}[t!]
\begin{center}
\includegraphics[]{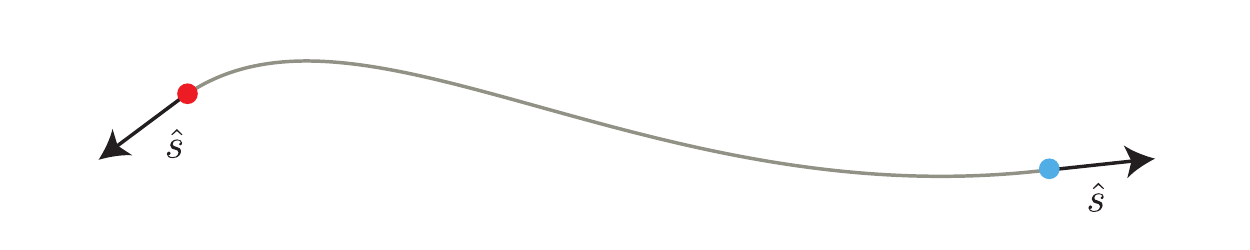}
\begin{minipage}[t]{5in}%
\caption[]{Definitions used in quirk equations of motion.}
\label{fig:Qeom}
\end{minipage}
\end{center}
\end{figure}

The action for a pair of heavy quarks connected by a gauge string
can be written
\beq
S = -\mq \sum_{i = 1}^2 \myint d\tau_i
- \si \myint dA
+ S_{\rm ext},
\eeq
where $d\tau_i$ is the proper length of the worldline
for quirk $i$, $dA$ is the proper area element of the
string worldsheet, 
and $S_{\rm ext}$ represents the effect of external forces.
Here $\si \sim \La^2$ is the string tension.
In the variation with respect to the quirk position, 
there is a surface term from the string action that generates
the string force on the quirks.
We therefore obtain the quirk equation of motion
\beq[releom]
\frac{\d}{\d t} \left( m \ga \vec{v}\gap \right)
= -\si \left[ \sqrt{1 - \vec{v}_{\perp}^{\gap\gap 2}}\,
\hat{s} 
+ \frac{v_{\parallel}}{\sqrt{1 - \vec{v}_{\perp}^{\gap\gap 2}}}\,
\vec{v}_{\perp} \right]
+ \vec{F}_{{\rm ext}},
\eeq
where $\vec{v}$ is the quirk velocity,
and $\vec{v}_\parallel$ and $\vec{v}_\perp$ are the components
of the quirk velocity parallel and perpendicular to the string:
\beq
\vec{v}_\parallel = (\vec{v} \cdot \hat{s}) \hat{s},
\qquad
\vec{v}_\perp = \vec{v} - \vec{v}_\parallel,
\eeq
where $\hat{s}$ is a unit vector along the string
pointing outward at the endpoints (see Fig.~\ref{fig:Qeom}).
The second term in brackets is similar to a Lorentz force,
and is required by relativistic invariance.

The gauge string is a dynamical object with its own
complicated equation of motion.
However, if the quirks have no further interactions after they
are produced (\eg\ with matter in the detector) then in the center
of mass frame the string remains straight.
Therefore, the only long-wavelength excitations of the string
arise from quirk interactions with matter.
If the string force is much larger than matter forces
\beq
F_{\rm ext} \ll \La^2,
\eeq
then we expect that the string will remain approximately
straight in the center of mass frame.
The maximum force from either ionization or nuclear
energy loss is of order $(100\eV)^2$, so the
straight-string approximation is guaranteed to hold only for
$\La \gg 100\eV$.
Note that this translates to $L \ll 10~\mbox{m}$, so all
but the longest strings of interest in colliders can be 
approximated as straight.
The full string dynamics is sufficiently complicated that
it would be useful to check this by direct simulation.

A potential concern is that interactions of the quirks with 
matter involve collisions with momentum transfer that may
be larger than $\La$.
For relativistic quirks, the energy and momentum transfer
in these processes is of order
\beq
\bal
\De p_{\rm ion} &\sim m_e \sim \mbox{MeV},
\\
\De p_{\rm nuc} &\sim \LQ \sim \mbox{GeV}.
\eal\eeq
We now ask whether this leads to the emission of infracolor glueballs.
The important point is that only the quirk has electromagnetic or
QCD interactions, so this energy and momentum transfer is to the quirk,
not the infracolor string, which is sterile under the standard model.
The change in the quirk velocity is of order
\beq
\De v \sim \frac{\De p}{\mq} \ll 1.
\eeq
This is a small perturbation as seen by the infracolor interactions,
and does not lead to the emission of an infracolor glueball.
This is very clear if we consider a heavy stable quark, which is surrounded
by QCD brown muck, but has no string attached.
In this case, the perturbation is equivalent to the quark remaining
at rest while the brown muck gets a velocity $\De v$ in the opposite
direction.
This transfers energy $\LQ \De v^2 \ll \LQ$ to the brown muck.
If this energy is smaller than the mass of the lightest hadron
that can be emitted (a pion in this case),
there is no transition and the process is elastic.
For quirks, the total mass of a long string may be much
larger than $\La$, but glueball emission is a local process
with a scale set by $\La^{-1}$.
We therefore expect hadron emission to be suppressed
as in the QCD case.

\section{Macroscopic Strings}
\label{sec:macrostrings}
We now consider strings with lengths longer than the tracking
resolution of a typical detector, very roughly
$L \gsim \mbox{mm}$.
In this case, the quirk and the antiquirk appear as separated
particles connected by a string.
Strings much longer than a detector size will not have observable
effects on the quirk trajectories, so we are considering
$\mbox{mm} \lsim L \lsim 10~\mbox{m}$
corresponding to
\beq
100\eV \lsim \La \lsim 10\keV
\eeq
for $\mq \sim \mbox{TeV}$.

\subsection{Anomalous Tracks}
\begin{figure}[t!]
\begin{center}
\includegraphics[]{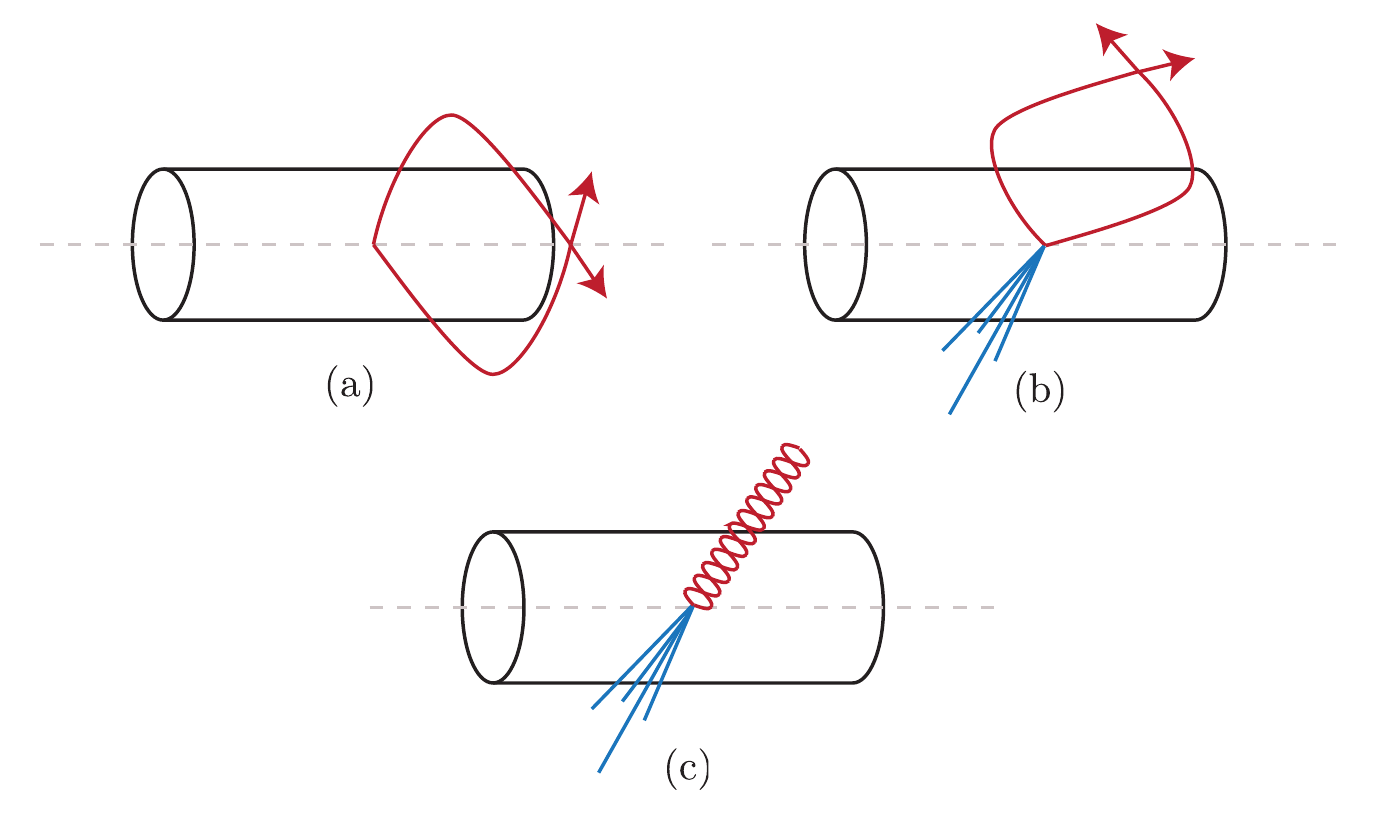}
\begin{minipage}[t]{5in}%
\caption[]{Anomalous tracks from quirks with macroscopic strings.}
\label{fig:qtrack}
\end{minipage}
\end{center}
\end{figure}

One obvious signature in this case is the anomalous quirk
tracks in the case where one or both quirks are electrically
charged.
Because the string tends to accelerate the quirks toward
each other, we can have events such as those depicted
schematically in Fig.~\ref{fig:qtrack}.
In these events, the curvature of the tracks
is qualitatively different from the curved track of a
particle in the magnetic field of the detector.
For example, a magnetic field along the beam direction
curves tracks only in the $r$-$\phi$ plane, while quirk
tracks generally have curvature in the $r$-$z$ plane.
Therefore, unambiguous observation of only a single event of 
this type is sufficient for discovery of macroscopic strings!

Do quirks annihilate when
the string force brings them back together?
For the case of macroscopic strings considered here, this is
highly suppressed by the fact that annihilation requires the
quirk to be in a state of relative angular momentum
$\ell \sim 1$, while interactions with matter change
the angular momentum by much larger amounts due to the
long lever arm.
Even a single ionization interaction gives
\beq
\De \ell \sim \De p L \sim m_e \frac{\La^2}{\mq} \sim
\left( \frac{\mq}{\mbox{TeV}} \right)
\left( \frac{\La}{\mbox{GeV}} \right)^{-2}.
\eeq
The infracolor ``brown muck'' surrounding the quirk has
a much larger cross section of order $\La^{-2}$,
and can therefore interact for angular momenta
$\ell \lsim \mq/\La$.
A single ionization interaction
changes the angular momentum more than this for
$\La\lsim\mbox{MeV}$.
We conclude that quirks with macroscopic strings do not
annihilate.

The difficulty in detecting quirks with macroscopic
strings is that triggers and track reconstruction algorithms
are designed for conventional tracks, and will likely miss these
events altogether.
Defining an efficient trigger for these events that has low
background from standard physics and instrumental noise is
worth further investigation.
A simpler strategy is to focus on events where the quirk pair
is produced in association with one or more hard jets or photons
(see Fig.~\ref{fig:qtrack}b and \ref{fig:qtrack}c).
Standard reconstruction algorithms will fail to reconstruct
the quirk tracks, resulting in missing $p_T$ balanced by
jets or photons.
If such events are discovered, careful examination of the
signal events in the missing $p_T$ direction can reveal the
presence of ``quirky'' tracks.

\subsection{Stopping Quirks}
Do quirks stop in the detector?
The stopping of heavy stable charged and/or strongly-interacting
particles has been extensively studied \cite{stopping}, with the
conclusion that typically a significant fraction do indeed stop
inside the detector.
For quirks there is an additional complication from the string
interaction.
In order for quirks to come to rest, they must become bound
to the lattice in the detector material.
If the string force is stronger than the 
forces that bind the quirks to the lattice, they will
continue to be dragged by the string.

We first consider possible final states of quirks bound to 
the lattice.
The binding mechanism depends on the standard model quantum
numbers of the quirks.
If quirks are electrically charged but uncolored,
they can be electronically bound
to the lattice similarly to ordinary nuclei.
This is particularly clear for positively charged quirks,
which can share a lattice site with an ordinary nucleus
since there is no constraint from the exclusion principle.
Negatively charged quirks will experience an electrical
potential with opposite sign, and it is reasonable to assume
that they will also find a stable local minimum.
If quirks are colored, they will form quirk hadrons whose
charge may change with
time because of inelastic strong interactions that change
the valence quark structure.
We expect quirk hadrons to bind efficiently with nuclei,
and these can also become
stuck in the lattice.

In all of these cases, the binding energy of the quirk
(or quirk-nucleus bound state) to the lattice is of
order eV, and the typical size of a potential well
is of order \AA.
(The binding energy is set by the electron mass, and is
independent of the mass of the heavy particle.
For example, the binding energies for heavy and light
nuclei are all several eV.)
Therefore, the force required to remove a bound quirk from the
lattice is of order
\beq
F_{\rm latt} \sim \frac{\mbox{eV}}{\mbox{\AA}} \sim (100\eV)^2.
\eeq
If the string force is larger than this, the lattice cannot
bind the quirk and it will not stop.

Even if $\La \ll 100\eV$, the string force gives the
quirks substantial kinetic energy, making it more unlikely for
them to bind to the lattice.
The binding energy of a nucleus in the lattice is of
order eV, and it is reasonable to assume that the lattice
cannot absorb energy larger than this without breaking.
Therefore, a quirk nucleus will not bind with
the lattice if its kinetic energy is large compared to eV.
This requires
\beq[betastoppingbound]
\be \lsim 10^{-6} \left( \frac{\mq}{\mbox{TeV}} \right)^{1/2}.
\eeq
For such small values of $\be$ ionization forces are
described by the theory of Fermi and Teller \cite{FT},
extended by Lindhard \cite{Lindhard}.
We have
\beq[slowionforce]
F_{\rm ion} \sim \La_0^2 \be,
\qquad
\La_0 \sim \mbox{keV}.
\eeq
The ionization force for different nuclei in the same material
vary over about an order of magnitude, suggesting an
uncertainty of an order of magnitude in $\La_0$.
Balancing this against the string force gives a terminal speed
\beq[termvel]
\be_* \sim \left( \frac{\La}{\mbox{keV}} \right)^2.
\eeq
Imposing \Eq{betastoppingbound} then gives
\beq
\La \lsim \mbox{eV} \left( \frac{\mq}{\mbox{TeV}} \right)^{1/4}.
\eeq
This bound is proportional to $\La_0$, so there is an uncertainty
of an order of magnitude in this estimate.
Despite this uncertainty, it
seems unlikely that quirks stop in the detector even
for the smallest values of $\La$ of interest.

If one quirk stops in the detector, the other will eventually
lose its kinetic energy and annihilate with it.
If both quirks stop, there is a string stretched between them.
This string can interact with strings of subsequently produced
quirks, producing even more bizarre events.
One can also imagine releasing such quirks by \eg\ melting
the material in which they are trapped, and looking for the subsequent
annihilation.
These are amusing possibilities that might be worth taking seriously
if more a detailed study indicates that large numbers of quirks
in fact stop in the detector.

\subsection{Quirk Annihilation}
Although the string force tends to prevent the quirks from
stopping, it also tends to bring them together to allow
them to annihilate.
In particular, sufficiently slow charged quirks (or quirk nuclei)
will reach the terminal speed given by \Eq{termvel}.
The subsequent motion of the quirks is damped,
so these quirks can find each other more efficiently.
Of course, this mechanism only works if both quirks are electrically
charged.

The linear form of the damping force holds up to
velocities of order $\al$.
For such small velocities
slow-moving colored quirks can bind with nuclei,
making them effectively
charged and subject to the mechanism considered here.
The maximum ionization force is of order $(100\eV)^2$,
so this mechanism only works for $\La \lsim 100\eV$.
Ionization energy loss is a good description as long as the
separation of the quirks is larger than atomic distances
of order $\mbox{\AA}$.
Even for distances smaller than \AA the energy loss
is more complicated, but we expect charged
quirks to exchange energy efficiently with electrons.
We therefore assume that they annihilate
on a time scale relevant for collider searches.

\begin{figure}[t!]
\begin{center}
\includegraphics[]{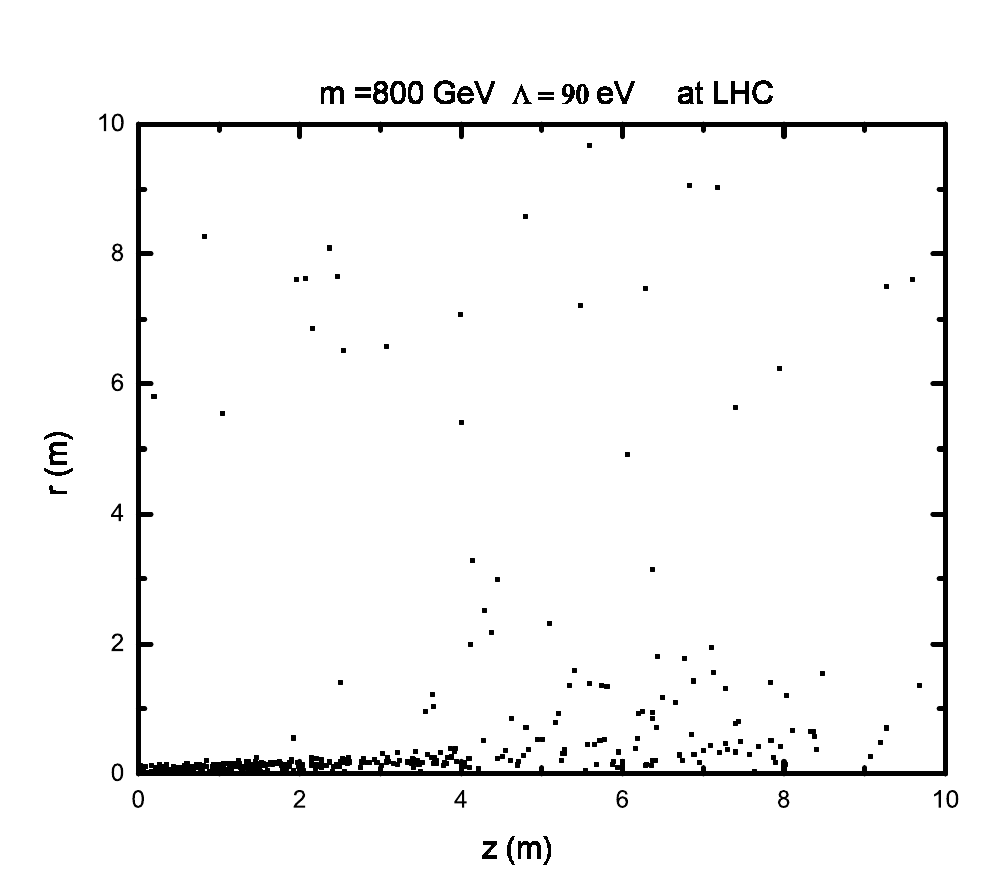}
\begin{minipage}[t]{5in}%
\caption[]{Results from a crude simulation of
position of quirk annihilation events relative to
collision point.}
\label{fig:reann}
\end{minipage}
\end{center}
\end{figure}

An interesting question is the
distribution of annihilation events in the detector.
We have made a crude simulation of this
using the straight string approximation.
We include ionization energy loss as a continuous force,
approximating the detector as solid iron.
We also include a crude approximation to nuclear energy
loss, although that does not really affect our results.
We assume that all quirks that reach terminal speed
and come close together in the detector annihilate sufficiently
rapidly to be seen.
An example of our results are shown in Fig.~\ref{fig:reann}.
Note that most of the annihilations take place near the beam.
This is easy to understand.
Most of the events where both quirks become damped
arise from events where quirks are produced nearly
back-to-back in the central region of the detector.
In such events the quirks will have speed less than
$\al$ at the turning points, and therefore they become
damped there.
Their subsequent motion is essentially constant velocity
toward each other, and they meet near the beam axis.

The distribution of annihilation events is very different
that of late-decaying particles stopped in the detector
\cite{stopping}.
More realistic simulations should be done to check the distribution
of these events.
Another difference is that most examples of late-decaying
particles that have been discussed in the literature
decay partly to missing energy while quirks will annihilate
to visible energy in most modes.

Another aspect of quirk annihilation that can in principle
give a signal is the ionization track of the damped
quirks before they re-annihilate.
The ionization is large compared to typical particles, but the
track is very slow ($\be \lsim \al$).
Presumably, it will therefore generate ``stub'' tracks in many
events that are triggered for other reasons, and these
stubs can in principle be connected.
Since these tracks lead to annihilation events
(assuming that the timescale for energy loss is sufficiently short)
so one can start looking for them there.

\begin{figure}[t!]
\begin{center}
\includegraphics[]{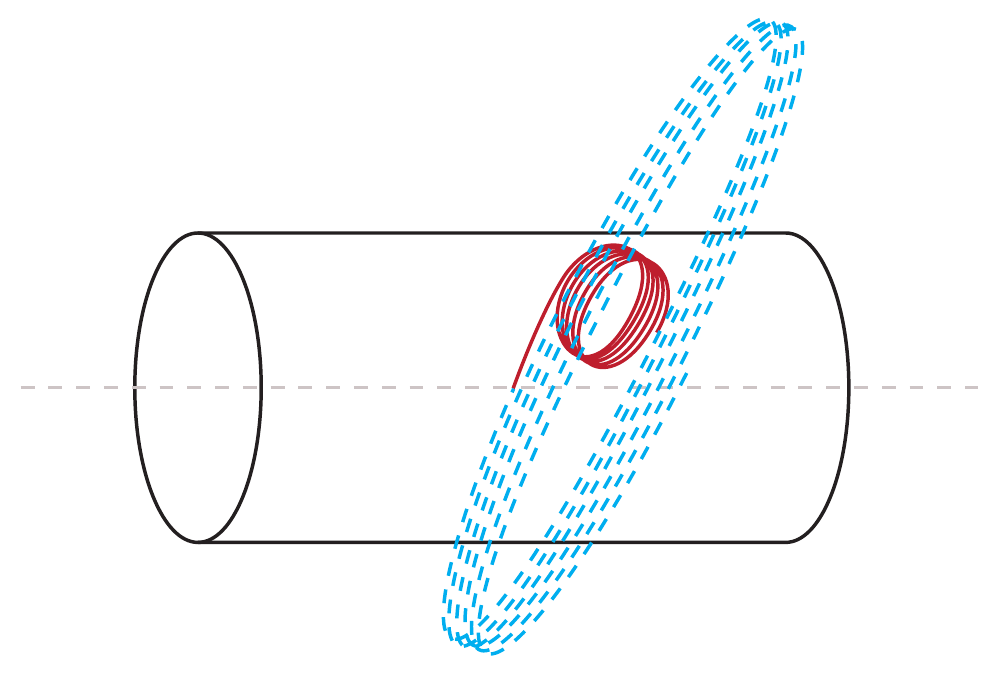}
\begin{minipage}[t]{5in}%
\caption[]{Highly exotic track resulting from an electrically charged
quirk (solid track) becoming damped in the detector, while its neutral
partner (dashed track) drives its motion.
The neutral track will be unobservable.}
\label{fig:bizarre}
\end{minipage}
\end{center}
\end{figure}

The previous discussion assumes that both quirks are
electrically charged, so that they both experience ionization
forces.
If the quirks are colored, their charge state may change
on a distance scale given by the nuclear mean free path
($\sim 10\cm$ in iron) complicating the phenomenology further.
One other case that bears mentioning is the case of
uncolored quirks where one is charged and the other is neutral,
\eg\ produced by $s$-channel $W$ exchange.
In this case, the charged quirk can become damped, while the
neutral quirk will not interact with matter.
In this case, the charged quirk will be driven by the
invisible neutral quirk which can have a much larger amplitude
of motion.
This can result in truly bizarre charged
tracks such as the one illustrated in Fig.~\ref{fig:bizarre}.
Since the damped quirk is moving very slowly ($\be \lsim \al$)
these events will be very difficult to detect.

\section{Prompt annihilation}
\label{sec:reann}
We now consider in more detail the question of quirk 
annihilation, which is very important for the phenomenology
of microscopic strings.
The momentum transfer in the annihilation process is of order
$\mq$, which means that the quirks must come within a distance
of order $\mq^{-1}$ in order to annihilate.
Equivalently, the cross section is dominated by
partial waves with relative angular momentum $\ell \lsim p/\mq \lsim 1$.
Because the maximum quirk separation $L$ is much
larger than the microscopic scales $\mq^{-1}$ and $\LQ^{-1}$,
there is a large lever arm with which interactions
with matter can change the angular momentum.
However, if the string is sufficiently short matter effects
are not important (we will be more precise about this below).
In this section we analyze annihilation of quirks
in the absence of matter effects.

A crucial question is the rate of transfer of energy and
angular momentum from the bound state.
An important feature is interactions of 
the non-perturbative ``brown muck'' surrounding the quirks,
from infracolor and/or QCD interactions.
The cross section for these interactions is much larger than the
hard annihilation of quirks, and may change the energy
and angular momentum of the system, 
thereby suppressing annihilation.
We also consider the effects of radiation as a mechanism
of losing energy and angular momentum, and we argue that this
is generally unimportant as a mechanism of energy loss.

\subsection{Quirkonium}
Quirk pairs produced near threshold can form a Coulomb-like
``quirkonium'' bound state.
Formation of such a low-lying bound state requires that the
quirk pair be produced near threshold, \ie
$|E - 2 \mq| \lsim \al^2 \mq$, where $\al$ is the infracolor
gauge coupling, or the QCD gauge coupling if the quirks are
colored and $\La < \La_{\rm QCD}$.
These bound states will annihilate promptly
into pairs of gluons or quarks (for colorful quirks).
This signal has been considered for stable gluinos
in \Ref{Cheung} where it was found to be less sensitive
than searches for unbound gluinos.
We expect the result for quirks to be qualitatively
similar, in that signals for highly excited quirks
($E - 2 \mq \gg \al^2 \mq$) will be more sensitive.

\subsection{Highly Excited Bound States}
We are interested in the majority of events that produce
quirks that are not close to threshold, \ie $E - 2\mq \sim \mq$.
As discussed in Section~\ref{sec:production}
quirk production is essentially perturbative,
so quirk pairs are produced in a state of relative angular
momentum $\ell \sim 1$.
The subsequent infracolor (and possibly QCD) ``hadronization''
stage does not strongly affect the energy and angular momentum
of the quirks, so the quirk pair still has $\ell \sim 1$
even when the quirks have large separation (\eg\ $r \gg \La^{-1}$).

If there are no further interactions that change the quirk angular
momentum, then the quirk pair will not have a well-defined
angular direction even if it is macroscopic!
If the quirks interact with matter (\eg\ in the detector)
their angular position will certainly be determined, but if
we are interested in cases with sufficiently large $\La$
then the typical length of a quirk string will be small
(\eg\ $\lsim \mbox{\AA}$) and matter interactions cannot
``measure'' the angular position.
In this case, a collider will create the quirk pair in a
``Schr\"odinger cat'' state with large-scale
($r \sim \mq / \La^2 \gg \La^{-1}$) quantum correlations.

Such a situation is not familiar in particle physics, and we
will  proceed cautiously.
In the end, many of the results can be understood from a simple
classical picture, but we will derive the results using
WKB wavefunctions to take into account the important
quantum-mechanical aspects of these states.

\subsection{Wavefunction Overlap}
\label{subsec:qm}
We now consider the probability that a highly excited
quirk pair can be found
sufficiently close together to re-annihilate.
This is a standard problem in quantum mechanics,
and we review it here to set the stage for the subsequent
discussion.
Highly excited states can be described using the WKB approximation.
The quirk annihilation probability
is proportional to the probability to find the
quirk pairs within a distance of order $\mq^{-1}$ of each other.
We will estimate this probability using non-relativistic quantum
mechanics and simple approximations that are sufficient
for our purposes.

We begin with the case $\ell = 0$.
We denote the radial Schr\"odinger wavefunction 
by $\psi(r)$ and define the
reduced wavefunction by
\beq
y(r) = \frac{\psi(r)}{\sqrt{4\pi}\, r}.
\eeq
This satisfies the normalization condition
\beq
1 = \int_0^r |y(r)|^2
\eeq
and the boundary condition 
\beq
y(0) = 0.
\eeq
The time-independent Schr\"odinger equation can then be written
\beq
y''(r) = - k^2(r),
\eeq
where
\beq
k(r) = \frac{\sqrt{2\muq}}{\hbar} \sqrt{K - V(r)}\,.
\eeq
Here $K = E - 2\muq$ is the kinetic energy,
and $\muq = \mq/2$ is the invariant mass of the reduced system.
We temporarily
keep $\hbar \ne 1$ to keep track of the classical limit.
We are interested in the case of a linear potential,
but we will see that the important results of this
section are independent of the details of the potential,
so we will keep it general.

We approximate the wavefunction of this system by
\beq
y(r) \simeq \frac{C}{\sqrt{k(r)}} \sin\left[
\int_0^r dr'\, k(r') \right]
\th(r_{\rm max} - r).
\eeq
where $r_{\rm max}$ is the classical turning point,
\ie
\beq
k(r_{\rm max}) = 0.
\eeq
For $r \ll r_{\rm max}$ this is the WKB wavefunction,
and has the correct boundary condition at $r = 0$.
The boundary condition at the classical turning point
is only crudely approximated, but this will not strongly
affect the probability to find the particle near the origin.
In this approximation, we can compute the normalization
constant as
\beq
1 = |C|^2 \int_0^{r_{\rm max}} dr\,
\frac{1}{k(r)} \sin^2 \left[ \int_0^r dr'\, k(r') \right].
\eeq
For highly-excited states, we are averaging over many
periods with a slowly-varying potential, so we can
replace $\sin^2$ by its average value $\frac 12$.
This gives
\beq
1 = \frac{\hbar |C|^2}{\sqrt{8\muq}} \int_0^{r_{\rm max}}
\frac{dr}{\sqrt{K - V(r)}}.
\eeq
This is directly related to the time for a classical trajectory
to go from $r = 0$ to $r = r_{\rm max}$:
\beq
T = \sqrt{\frac{\muq}{2}} \int_0^{r_{\rm max}} 
\frac{dr}{\sqrt{K - V(r)}}.
\eeq
For a linear potential
\beq
V(r) = \si r
\eeq
we have
\beq[Tlinearpot]
T = \frac{\sqrt{2 \muq K}}{\si}.
\eeq
However, it is more insightful to leave the results in terms of
$T$ as we will see.
We therefore have
\beq[CTresult]
|C|^2 = \frac{2\muq}{\hbar T}.
\eeq

We now estimate the probability
to find the quirks within a distance $r_0$ of each other:
\beq
\mbox{Prob}(r \le r_0) = \int_0^{r_0} dr\, |y(r)|^2.
\eeq
Near the origin, the wavefunction is oscillating with
the de Broglie wavelength
\beq
\la_0 = \frac{2\pi\hbar}{\sqrt{2 \muq K}}.
\eeq
For $r_0 \gg \la_0$ the integral averages
over many periods, and we can again replace the $\sin^2$ term
by its average value of $\frac 12$:
\beq[classtimeresult]
\mbox{Prob}(r \le r_0) \simeq \sfrac 12 |C|^2 \int_0^{r_0} \frac{1}{k(r)} 
= \frac{\De t}{T},
\eeq
where 
\beq
\De t = \sqrt{\frac{\muq}{2}} \int_0^{r_0} \frac{dr}{\sqrt{K - V(r)}}
\eeq
is the classical time to go from $r = 0$ to $r = r_0$ and
we have used \Eq{CTresult} to eliminate $C$.
This is the result familiar from quantum mechanics textbooks
that in a highly excited state
the probability to find a particle at the origin is 
proportional to the fraction of time that a classical orbit
spends there.

In the opposite limit
$r_0 \ll \la_0$ we can use the approximation
\beq
\sin\left[ \int_0^r dr'\,
k(r') \right]
\simeq \sin k_0 r \simeq k_0 r,
\eeq
where $k_0 = 2\pi / \la_0$.
Since the wavefunction at the origin is
\beq
\psi(0) = C \sqrt{\frac{k_0}{4\pi}}
\eeq
this gives
\beq[wavefcnoriginresult]
\mbox{Prob}(r \le r_0) = |C|^2 \int_0^{r_0} k_0 r^2
= |\psi(0)|^2 V,
\eeq
where $V = \frac 43 \pi r_0^3$ is the volume of the region
of interest.
This is the result familiar from positronium and quarkonium
physics that the $\ell = 0$
annihilation probability is proportional to the wavefunction
at the origin.

The results \Eqs{wavefcnoriginresult} and \eq{classtimeresult}
are very different parametrically.
For $r_0 \gg \la_0$ the result is classical
and therefore independent of $\hbar$, which is not the
case for $r_0 \ll \la_0$.
Also, for $r_0 \gg \la_0$ the probability goes
as $r_0$ (since $\De t \sim r_0 / v$ where $v$ is the 
velocity of the classical trajectory near the origin),
while for $r_0 \ll \la_0$ the probability goes as $r_0^3$.

We will be mainly interested in the limit $K \ll \muq$,
where $r_0 \ll \la_0$.
From \Eqs{wavefcnoriginresult} and \eq{CTresult} we have
\beq[wavefunctionoriginapprox]
|\psi(0)|^2 = \frac{\sqrt{2 \muq^3 K}}{2\pi T}
= \frac{\mq^2 \be}{4\pi T},
\eeq
where we have expressed the result in terms of the physical
quirk mass and the velocity
of a single quirk at production, given by
\beq
\be = \left( \frac{2K}{\mq} \right)^{1/2}.
\eeq
For the majority of events $\be \sim 1$ and
therefore $|\psi(0)|^2 \sim T^{-1}$.
As we will see, this means that for highly excited quirk bound states
there is a definite annihilation probability per
classical crossing.
\Eq{Tlinearpot} shows that
$T \propto \be$, so the probability is nonzero at threshold.

It is straightforward to 
include the effects of nonzero orbital angular momentum.
The angular momentum barrier means that classically the
particles have a distance of closest approach given by
\beq
r_{\rm min} = \frac{\ell}{\sqrt{2 \mq K}} \sim \ell \la_0,
\eeq
where $\ell$ is the angular momentum.
(We are setting $\hbar = 1$ again.)
For $r \le r_0 \ll \la_0 \lsim r_{\rm min}$
we can use the approximation
\beq
y_{\ell}(r) \simeq C_\ell r^{\ell + 1}.
\eeq
We can determine the coefficients $C_{\ell}$ by matching onto
the wavefunction for $r \gsim r_{\rm min}$.
Since $r_{\rm min} \gsim \la_0$, the sine function in the
wavefunction is of order 1, and we have
\beq
y_\ell(r_{\rm min}) \sim \frac{C}{k(r_{\rm min})} 
\sim C_\ell r_{\rm min}^{2\ell + 1}.
\eeq
With this approximation we obtain
\beq
\mbox{Prob}(r \le r_0) \sim \frac{1}{2\ell + 3}
\left( \frac{\muq}{K} \right)^{1/2}
\frac{r_0^{2\ell + 3}}{r_{\rm min}^{2\ell + 2}}\,
\frac{1}{T}.
\eeq
For $r_0 \sim \mq^{-1}$ this is suppressed compared to
the $\ell = 0$ case by
\beq
\frac{\mbox{Prob}_{\ell \ne 0}(r \le \mq^{-1})}
{\mbox{Prob}_{\ell = 0}(r \le \mq^{-1})}
\sim \frac{1}{\ell}
\left( \frac{\be}{\ell} \right)^{\ell + 1}.
\eeq
This suppression means
that annihilation is dominated by small $\ell$.

\subsection{Annihilation Rates}
\label{subsec:reannrate}
We now use the results above to compute the quirk annihilation
rates.
For now we neglect the effects of interactions with matter,
non-perturbative interactions, and radiation.
We work in the highly excited regime
$\al^2 \mq \ll K \ll \mq$.
Our results should be approximately valid for $K \lsim \mq$,
the regime where the majority of quirk pairs are produced.
In this regime, the state is sufficiently excited to use
the WKB approximation of the previous subsection,
but the de Broglie wavelength of the quirk is larger
than the distance $r_0 \sim \mq^{-1}$ over which the
annihilation takes place.
As we reviewed above,
the probability to find the quirks near the origin is
dominated by the $\ell = 0$ partial wave, and is
proportional to $|\psi(0)|^2$.
The density of particles within the range of the annihilation
cross section is therefore $|\psi(0)|^2$, and the
annihilation rate is
\beq
\Ga = |\psi(0)|^2 \si v_{\rm rel},
\eeq
where $\si$ is the annihilation cross section and $v_{\rm rel}$
is the relative velocity of the quirks in their center
of mass frame.
The wavefunction at the origin is given by 
\Eq{wavefunctionoriginapprox}, and is proportional to 
$1/T$, where $T$ is the classical time for the quirks to
go from $r = 0$ to $r = r_{\rm max}$.
The annihilation probability per classical crossing is therefore
\beq[probannihilation]
P = 2 T \Ga = \frac{\mq^2\be}{2\pi} \si v_{\rm rel}.
\eeq
Note that all dependence on the potential has dropped out,
so the time scale for annihilation is set
by the classical crossing time.

The annihilation cross sections can be computed perturbatively.
To get numerical factors right, note that
the spin of the quirks is not correlated, and so we must
average over intial spins.
Similarly, the QCD color of colored quirk pairs is uncorrelated,
so we average over quirk colors.
On the other hand, quirks are in an infracolor
singlet state because they are connected by an infracolor
string.

For quirks carrying QCD color, we then have
\beq[QQgg]
\Ga(Q\bar{Q} \to gg) &= \frac{16 N_{\rm IC}}{27}\,
\frac{2\pi \al_3^2}{\mq^2} |\psi(0)|^2,
\\
\eql{QQqq}
\Ga(Q\bar{Q} \to u\bar{u}) &= \frac{2 N_{\rm IC}}{9}\,
\frac{\pi \al_3^2}{\mq^2} |\psi(0)|^2,
\eeq
where $\si v_{\rm rel}$ has been replaced by its threshold value.
The QED annihilation processes are
\beq[QQgaga]
\Ga(Q\bar{Q} \to \ga\ga) &= \frac{N_{\rm IC} e_Q^4}{N_{\rm C}} \,
\frac{2\pi \al^2}{\mq^2} |\psi(0)|^2,
\\
\eql{QQee}
\Ga(Q\bar{Q} \to \ga^* \to e^+ e^-) &= \frac{N_{\rm IC} e_Q^2}{N_{\rm C}}\,
\frac{\pi \al^2}{\mq^2} |\psi(0)|^2,
\eeq
where $e_Q$ is the electric charge of the quirk
and $N_{\rm C}$ is the number of QCD colors of the quirk
(so $N_{\rm C} = 1$ if the quirks are color singlets).
We have neglected the contribution from $Z$ boson exchange,
which gives a small correction.
There are similar expressions for annihilation through a
$W$ in the case where the electric charge of the quirks 
differs by one unit.

For quirks carrying both QCD color and electric charge
there is a potentially interesting
mixed annihilation to gluons and photons with rate
\beq[QQgag]
\Ga(Q\bar{Q} \to g \ga) &= 
\frac{4 N_{\rm IC} e_Q^2}{9}\,
\frac{2\pi \al \al_3}{m_Q^2} |\psi(0)|^2.
\eeq
This motivates searches for photon-jet resonances at colliders.

There is also annihilation to infracolor gluons, which gives 
\beq[QQinv]
\Ga(Q\bar{Q} \to \mbox{infracolor})
= \frac{N_{\rm IC}^2 - 1}{4 N_{\rm IC} N_{\rm C}}\,
\frac{2\pi \al_{\rm IC}^2}{\mq^2}
|\psi(0)|^2.
\eeq
For $\La \lsim 10\GeV$, the infracolor glueballs are
stable on collider scales and this is an invisible decay.
The gauge couplings in \Eqs{QQgg}--\eq{QQinv} are to be evaluated at
a renormalization scale $\mq$.
For the infracolor coupling, we approximate
\beq
\al_{\rm IC}(\mq)
\simeq \frac{6\pi}{11 N_{\rm IC} \ln 4\mq/\La_{\rm IC}}.
\eeq
Here we estimate the scale where the perturbative
coupling blows up as $\La_{\rm IC}/4$.
(In QCD, this scale is $\simeq 250\MeV$, while the scale
of strong interactions is $\simeq \mbox{GeV}$.)

The annihilation probability per classical crossing time
is important in comparing the annihilation rate with other energy loss
mechanisms.
For colored quirks,
the annihilation into quarks and gluons dominates.
The probability of annihilation per classical crossing is
\beq[PannQCD]
P_{\rm QCD} = 2 T \Ga_{\rm QCD} = 
\frac{32 N_{\rm IC}}{27}\,
\al_3^2 \be 
\sim \frac{1}{40},
\eeq
where we have assumed that the quirks can annihilate into all
6 quark flavors, and we assumed
$N_{\rm IC} = 3$ and $\be \sim 1$.
For uncolored quirks with opposite charge,
the probability per classical crossing
to annihilate is
\beq[PannQED]
P_{\rm QED} = N_{\rm IC} (6 e_Q^2 + e_Q^4) \al^2
\be 
\sim \frac{1}{780},
\eeq
where we have used $e_Q = 1$, $N_{\rm IC} = 3$
for the numerical estimate.
The annihilation probability per classical crossing
to annihilate to infracolor gluons is
\beq[Panninv]
P_{\rm inv} = \frac{N_{\rm IC}^2 - 1}{2 N_{\rm IC}}
\al_{\rm IC}^2 
\be 
\sim \frac{1}{160} \, \frac{1}{[1 - 0.12 \ln(\La_{\rm IC} / \mbox{GeV})]^2},
\eeq
where we assume uncolored quirks with $N_{\rm IC} = 3$
and $\mq \sim \mbox{TeV}$.
We see that the annihilation into visible final states
is significant even for uncolored states with large values of $\La$.
These results will be useful in assessing the probability that
quirks undergo prompt annihilation.

Another important quantity is the branching ratio
for colored quirks to annihilate to leptons and photons.
We have
\beq[annleptonbr]
\frac{\Ga(Q\bar{Q} \to \mu^+ \mu^-)}
{\Ga(Q\bar{Q} \to jj)}
&= \frac{9 }{68}\, \frac{e_Q^2 \al^2}{\al_3^2}
\simeq 1.4 \times 10^{-4},
\\
\eql{annphotonbr}
\frac{\Ga(Q\bar{Q} \to \ga\ga)}
{\Ga(Q\bar{Q} \to jj)}
&= \frac{9 }{34} \, \frac{e_Q^4 \al^2}{\al_3^2}
\simeq 3 \times 10^{-5},
\eeq
where we have taken $e_Q = \frac 13$ for the numerical values.
These branching ratios are discouragingly small.%
\footnote{These branching fractions are lower than the corresponding
ones for the Upsilon decays mainly because the initial state
is not a color singlet.
This opens additional colored channels
and enhances the strong decay rate.}
On the other hand, the branching ratio to photon plus jets is
\beq
\frac{\Ga(Q\bar{Q} \to g\ga)}
{\Ga(Q\bar{Q} \to jj)}
= \frac{6}{17} \, \frac{e_Q^2 \al}{\al_3}
\sim 3 \times 10^{-3},
\eeq
where we again take $e_Q = \frac 13$.
This is somewhat more encouraging.

The accuracy of the estimates above can easily be
improved by incorporating the behavior of the cross
section at threshold and relativistic effects.
We leave this to future work.

The annihilation probabilities computed above are
relevant when there are no interactions that can change
the angular momentum of the quirks.
We now consider these interactions to see whether they are
in fact negligible.

\subsection{Non-perturbative QCD Interactions}
We now consider the effects of non-perturbative QCD interactions
on colored quirk annihilation.
Colored quirks are surrounded by a cloud of non-perturbative QCD
``brown muck'' with size
$R_{\rm had} \sim \LQ^{-1}$.
Interactions between the brown muck of the quirks are
important because they have a larger
cross section than the hard annihilation processes considered
previously.
Although they do not result in the annihilation of the quirk pair,
they can change the angular momentum of the quirk pair and
affect the probability for hard annihilation.

We assume that $L \gg R_{\rm had}$,
so that we can treat the quirk hadrons as well-separated
particles moving under the influence of an
infracolor string.
This requires
\beq
E \gsim \frac{\La^2}{\LQ}.
\eeq
For example, for $\La \sim \mbox{GeV}$ this requires only
$E \gsim \mbox{GeV}$, a mild requirement for hard production.
For $\La \gsim 10\GeV$, this starts to be a significant
constraint, and our results will be qualitatively
reliable at best.

The brown muck will interact only when the
quirks come within a distance of order $R_{\rm had}$.
We are interested in inelastic processes (\eg\ pion emission)
that can change the energy and angular momentum of the bound
state.
The typical energy transfer from the bound state can be estimated
from
\beq
\De E \sim F \De r \sim \LQ^2 R_{\rm had} \sim \LQ.
\eeq
The momentum transfer can in principle be larger if the quirks
are moving slowly:
\beq
\De p \sim F \De t \sim \LQ^2 \frac{R_{\rm had}}{v} \sim \frac{\LQ}{v}.
\eeq
However, for pion emission $\De p \sim \De E$, so energy
and momentum transfer are of order $\LQ$.
We expect such processes to have a geometric cross section
of order $\pi R_{\rm had}^2$, since there is no
small parameter suppressing the interaction probability.%
\footnote{If the number of QCD colors is regarded as a large
parameter, the probability for interaction is of order $1/N_{\rm C}$.
We will neglect large $N_{\rm C}$ effects in the following.}
Note that the large kinetic energy carried by the quirk is not
transfered in the interaction, and does not suppress the
interaction probability.

A geometrical cross section is equivalent to saturating
unitarity for all partial waves up to
\beq
\ell_{\rm max} \sim \mq v R_{\rm had} \sim (\mq E)^{1/2} R_{\rm had}.
\eeq
Unless we are very close to threshold we have $\ell_{\rm max} \gg 1$,
and so the partial wave cross section will approximately saturate
unitarity unless the angular momentum is very large.
This means that the interaction will take place with of order
unit probability whenever the quirks come within a distance
of order $R_{\rm had}$ or less.
We therefore expect that a brown muck interaction transfering
energy and momentum of order $\LQ$ will take place roughly
once every crossing time, as long as $\ell \lsim \ell_{\rm max}$.

We can understand this result using a simple quantum-mechanical model.
We model the brown muck as a particle (a constituent quark)
of mass $\sim \LQ$ bound to each quirk by a potential
that represents the
effects of the QCD interactions.
The wavefunction for the system is then a function of the relative
coordinate of the quirks $\vec{r}$ and the coordinates 
of the constituent quarks relative to the associated quirk
$\vec{\rho}_{1,2}$.
The wavefunction is assumed to take the approximate form
\beq
\Psi(\vec{r}, \vec{\rho}_1, \vec{\rho}_2)
\sim \psi(r) \chi(\rho_1) \chi(\rho_2),
\eeq
where $\psi(r)$ is the quirk wavefunction,
and $\chi(\rho)$ is the consituent quark wavefunction.
This factorized form is justified for $r \gg R_{\rm had}$ where
the quirks are well separated.
The constituent quark wavefunction $\chi(\rho)$ is nonzero only
for $\rho \lsim R_{\rm had}$.
It is convenient to normalize it so that
\beq
\myint d^3 \rho\, | \chi(\rho) |^2 \sim R_{\rm had}^3.
\eeq
We then have (see Subsection~\ref{subsec:qm})
\beq
\psi(r) \simeq \frac{C'}{r}\, \frac{1}{\sqrt{k(r)}}
\sin \left[ \int_0^r dr'\, k(r') \right].
\eeq
We are using an $\ell = 0$ wavefunction, which will have the right
qualitative behavior as long as $\ell \ll \ell_{\rm max}$.
Normalizing the wavefunction gives
\beq
|C'|^2 \sim \frac{m_Q}{\hbar R_{\rm had}^6 T},
\eeq
where $T$ is the classical crossing time.
Here we make the same approximations as previously for the quirk
wavefunction.
We can then compute the probability that the brown muck particles
are within a distance $R_{\rm had}$ from each other:
\beq
\!\!\!\!\!\!\!
\mbox{Prob}(|\vec{r} + \vec{\rho}_1 - \vec{\rho}_2| \le R_{\rm had})
&= \frac{|C'|^2}{2 k_0}
\myint d^3 r\, \frac{1}{r^2} 
\myint d^3\! \rho_1\,  |\chi(\rho_1)|^2
\myint d^3\! \rho_2\, |\chi(\rho_2)|^2
\nonumber\\
& \qquad\qquad\qquad
\times \th(R_{\rm had} - |\vec{r} + \vec{\rho}_1 - \vec{\rho}_2|)
\\
&\sim |C'|^2 \frac{\hbar R_{\rm had}^7}{(\mq E)^{1/2}}
\sim \frac{R_{\rm had} / v}{T},
\eeq
where $v \sim (E / \mq)^{1/2}$ is the classical quirk velocity
at the origin and we have again 
used the fact that the Compton wavelength of the the heavy
quirk is much smaller than $R_{\rm had}$.
This is the fraction of the time that the quirks are within
a distance $R_{\rm had}$.
To find the reaction rate, we must find the density of incident
particles over the range of the interaction.
This is
\beq
\rho \sim \frac{R_{\rm had}/v}{T} \frac{1}{R_{\rm had}^3}
\sim \frac{1}{R_{\rm had}^2 v T},
\eeq
so the reaction rate is
\beq
\Ga \sim \rho v \si \sim \frac{1}{R_{\rm had}^2 v T}
v  R_{\rm had}^2
\sim \frac{1}{T}.
\eeq
We are again led to the conclusion that these interactions
occur roughly once per classical crossing.

One important effect of these interactions is that
it changes the angular momentum state of the quirk pair.
The quirks are produced in a state with angular momentm $\ell \sim 1$,
\ie\ a highly spherical quantum state in which the angular position
of the quirks has nearly maximal uncertainty.
The hadrons that are emitted eventually interact with matter
far from the detector, and therefore can be thought of as having a definite
direction.
The fact that the angular momentum state of the hadrons is entangled
with that of the quirk pair means that this reduces the quantum
uncertainty in the the angular direction of the quirk pair.
In the traditional textbook language of quantum mechanics, the angular
position of the quirks gradually becomes ``measured'' by the
repeated ``measurement'' of the pion angular positions.%
\footnote{We are neglecting possible interactions of
the quirks with matter, which would directly ``measure'' the
angular position of the quirks.
This discussion is therefore applicable to the case where the 
string is sufficiently short that matter interactions are
unimportant.}
A proper treatment of this process using the ideas of quantum
decoherence is beyond the scope of the present work, and we will
only make some simple estimates here.

The angular momentum transfered to the emitted hadrons
in a single brown muck interaction is of order
\beq
\De \ell \sim R_{\rm had} \De p \sim 1.
\eeq
Assuming that the interaction is equally likely to raise
or lower the angular momentum, we have
$\ell \sim \sqrt{N}$ after $N$ such interactions.%
\footnote{The angular momentum is positive semi-definite,
but this is taken care of in the random walk by simply
identifying $\pm\ell$.}
The hard annihilation cross section falls rapidly for
$\ell \gsim \mbox{few}$,
so there is a competition between hard
annihilation, which wants to eliminate the bound state in
a small number of classical crossing times,
and the non-perturbative QCD interactions, which tend
to increase the average angular momentum, and therefore suppress
hard annihilation.

We can illustrate these points with a simple 
quantum mechanical toy model.
We work in 2 dimensions,
where the angular momentum eigenstates are simply
$e^{im\th}$, where $\th$ is the polar angle and $m$ is an integer.
We can simplify the model further by restricting the particles
to a circle, so there is no radial wavefunction to worry about.
We assume that there is a process by which a ``quirk bound state'' in an
angular momentum $m$ state emits a ``pion'' that also lives
on the circle.
The 1-particle wavefunction therefore makes a transition to a
2-particle wavefunction
\beq
e^{im\th} \to a_{0} e^{im\th}
+ a_{1} e^{i(m + 1)\th} e^{-i\th'} + 
a_{-1} e^{i(m - 1)\th} e^{i\th'} 
+ \cdots.
\eeq
Here $\th'$ is the angular coordinate of the emitted pion.
The transition conserves angular momentum since
$L = -i( \d_\th + \d_{\th'})$.
The amplitudes $a_0, a_{\pm 1}, \ldots$ can depend on $m$,
but we make the simplifying assumption that they are independent
of $m$.
We assume that $a_n$ is significant for $n \sim 1$, so it
is sufficient to consider $a_0$ and $a_{\pm 1}$.
Symmetry under $\th \to -\th$ then implies that $a_{1} = a_{-1}$,
and we have simply
\beq[toytrans]
\psi(\th) \to \psi'(\th, \th')
= \bigl[ a_0 + 2 a_1 \cos(\th - \th') + \cdots \bigr]
\psi(\th).
\eeq
We can choose $a_0$ real without loss of generality.
We assume that the pion emission is peaked at $\th = \th'$,
so that $a_1$ is mostly real.
We now imagine that the bound state repeatedly emits pions,
and the angular position of the pions is measured.
This corresponds to making the transition \Eq{toytrans}
and then fixing $\th'$ by picking a value of $\th'$ according
to the probability distribution
\beq
P(\th') = \myint d\th\, |\psi'(\th, \th')|^2.
\eeq
Picking $\th'$ in this way then gives a new wavefunction
that depends only on $\th$, which can then undergo further
transitions.

\begin{figure}[t!]
\begin{center}
\includegraphics[]{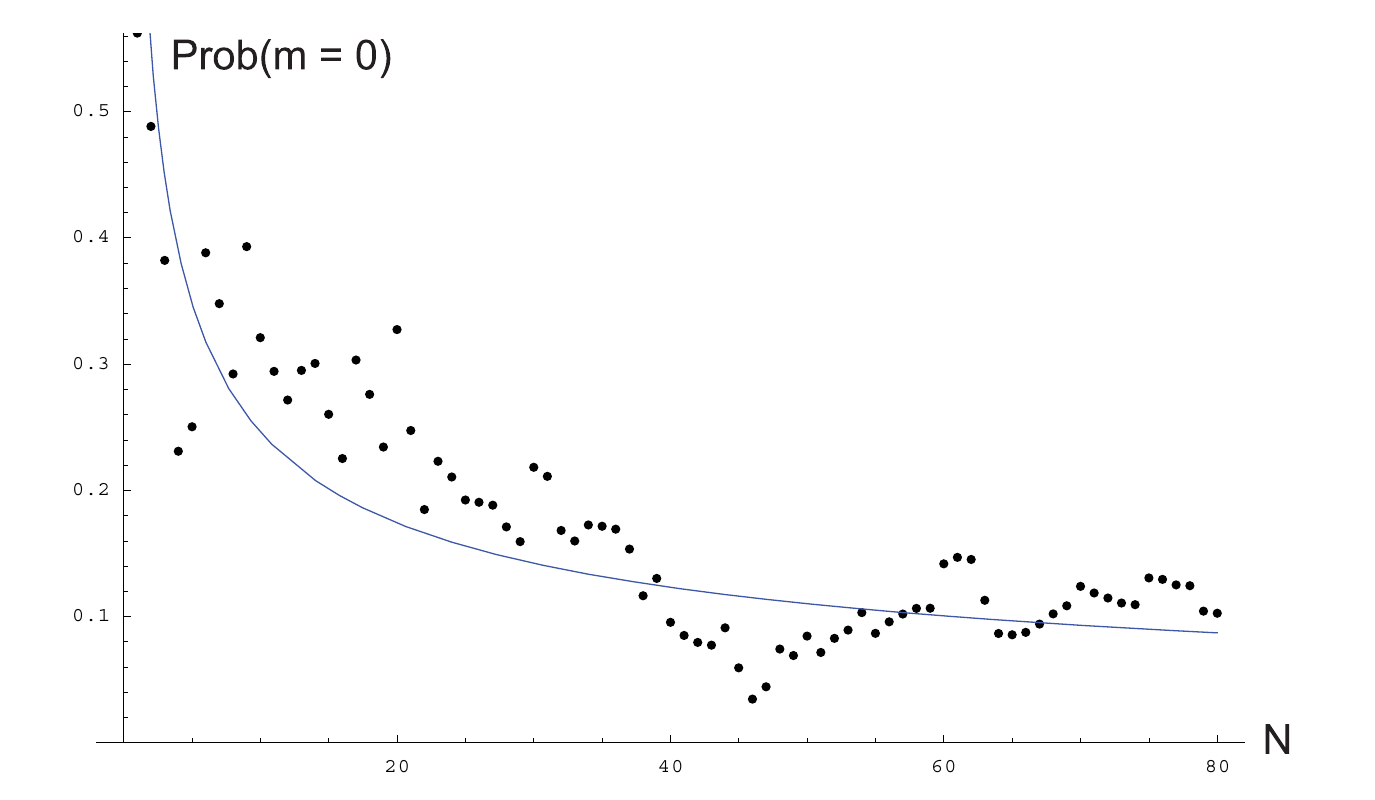}
\begin{minipage}[t]{5in}%
\caption[]{Monte Carlo simulation of toy model of angular
decoherence.
The plot shows the probability to find the system in
the $m = 0$ angular momentum state after $N$
interactions.
The curve is the fit to $\mbox{constant}/\sqrt{N}$.}
\label{fig:angdec}
\end{minipage}
\end{center}
\end{figure}

This simple toy model captures the basic quantum kinematics
of the problem we care about.
For example,
we can easily see how repeated transitions of the form \Eq{toytrans}
make the angular position more well-determined.
The value of $\th'$ is correlated with the direction
of $\th$, so this tends to make the peak more pronounced.
The quantity of most interest to us is the probability to find
the bound state in an $m = 0$ state after $N$ transtions.
This probability is expected to decrease as $\sim 1/\sqrt{N}$,
since each transition changes the maximum angular momentum by $\pm 1$.
This is born out by Monte Carlo simulation of this model
(See Fig.~\ref{fig:angdec}).
Although this toy model is a drastic simplification of the
system of interest, it illustrates that the expected behavior
does arise from quantum mechanics.
We therefore expect the same behavior in the realistic system.

\subsection{Hadronic Fireballs?}
The arguments above suggest that a significant fraction
of colored quirk pairs lose most of their energy to
emission of QCD hadrons.
This requires that the quirks do not annihilate for
a number of crossings of order $\mq / \LQ \sim 10^3$.
The non-perturbative QCD interactions remain effective up to very
large angular momenta, of order $\ell_{\rm max} \sim \mq/\LQ \sim 10^3$,
which takes would take of order
$10^6$ crossing times to reach according to the random-walk picture.
In the meantime, each non-perturbative QCD interaction results in
the emission of one (or several) hadrons with total energy 
$\sim \LQ \sim \mbox{GeV}$.
This means that the kinetic energy of the bound state
($K \sim \mq \sim \mbox{TeV}$)
is rapidly converted to $\sim 10^3$ hadrons with energy $\sim\mbox{GeV}$
each: a hadronic ``fireball.''

We can obtain a simple estimate of the fraction of events of this type
by assuming that the quirk survival probability
at the $n^{\rm th}$ crossing is $(1 - P / \sqrt{n})$,
where $P \sim 1/360$ is the $s$-wave annihilation
probability.
The probability to survive for $10^3$ crossings is then approximately
$85\%$.

When the quirks finally annihilate, they are essentially at rest
in their center of mass frame, so the annihilation products appear
as a narrow resonance with mass $2\mq$.
The intrinsic width will be due to the fact that the final
annihilation will take place from a distribution of low-lying
Coulombic ``quirkonium'' states.
The width will therefore be of order the spacing of
low-lying Coulombic energy levels, given by
\beq[energylevelCoulcolq]
\De E \sim \al_{\rm IC}^2(\mq) \frac{\mq}{2}
\sim 3\GeV \left(\ln 
\frac{m_Q / \mbox{TeV}}{\La / \mbox{GeV}}\right)^{-2}.
\eeq
This also sets the scale for the energy emission during
the final stages of the decay, which we see is only
slightly larger than the QCD scale.

The time for this process is set by the classical crossing
time and the number of interactions required to lose the
kinetic energy:
\beq[brownmuckdisplace]
c\tau \sim \frac{\mq}{\LQ} \frac{\mq}{\La^2}
\sim 10^{-2}\cm
\left( \frac{\La}{\mbox{MeV}} \right)^{-2}
\left( \frac{\mq}{\mbox{TeV}} \right)^2.
\eeq
We see that the decay may have a displaced vertex
for smaller values of $\La$.

The dominant decay will be to two jets, which may be a difficult
signal due to large backgrounds.
The decay to leptons or photons has a suppressed branching
ratio, but offers
a cleaner signal that may be easier to look for.
If energy loss due to QCD interactions is efficient,
the final hard annihilation of the quirks will be from
a Coulomb-like state that is color and infracolor singlet.
This means that there are fewer colored channels
compared to the excited annihilation computed in
Subsection~\ref{subsec:reannrate}.
Assuming $s$-wave annihilation we find
\beq
\frac{\Ga(Q\bar{Q} \to \mu^+ \mu^-)}
{\Ga(Q\bar{Q} \to \mbox{jets})}
&= 18 \frac{e_Q^2 \al^2}{\al_3^2} \simeq 2 \times 10^{-2},
\\
\frac{\Ga(Q\bar{Q} \to \ga\ga)}
{\Ga(Q\bar{Q} \to \mbox{jets})}
&= 36 \frac{e_Q^4 \al^2}{\al_3^2} \simeq 4 \times 10^{-3},
\eeq
for $e_Q = \frac 13$.
This looks very promising.
The decay to $g \ga$ is absent, although there is a suppressed
decay mode $gg\ga$.

Since the quirks lose all their kinetic energy before
decaying in these events, the decay products will have
an invariant mass very close to $2\mq$.
The intrinsic width will be due to the fact that the final
annihilation will take place from a distribution of low-lying
Coulombic states.
These have very small energy differences of order $\De E$
(see \Eq{energylevelCoulcolq}),
so the intrinsic width of the resonance is very small.

\begin{figure}[t!]
\begin{center}
\includegraphics[]{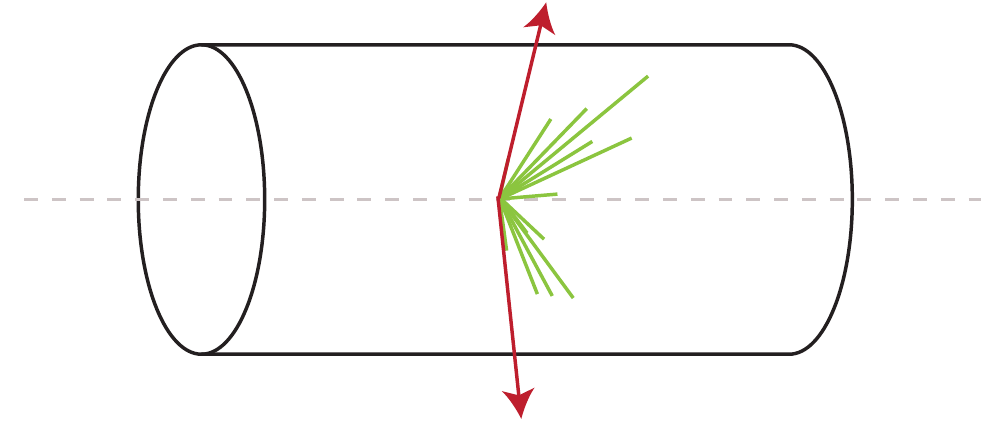}
\begin{minipage}[t]{5in}%
\caption[]{Schematic depiction of hadronic
fireball and hard annihilation into muons.
Note that the the asymmetry of the muons and
the fireball are in the same direction.}
\label{fig:fireball}
\end{minipage}
\end{center}
\end{figure}

Can we hope to see the hadronic fireballs associated with
these decays?
Most of the hadrons are expected to be pions.
Muons from charged pion decays
will be difficult to detect because they are highly curved
in the magnetic field of the detector.
Neutral pions decay to photons, which may be more promising to
detect.
The angular distribution of the fireball may aid in distinguishing
it from background.
Due to the angular decoherence, the quirk pair acquires an angular
position in the center of mass frame.
We expect that hadron emission
is peaked in the direction of the quirk motion,
resulting in a doubly-peaked pattern in the center of mass frame.
Note that the quirk annihilation is dominantly $s$-wave,
and so the direction of the annihilation products is not
correlated with the direction of the original quirk motion.
This means that the fireball generally does not line up
with the annihilation products.
Furthermore, the longitudinal boost of the center of mass
system will push both the fireball and the hard
annihilation products in the same direction.
This is illustrated in Fig.~\ref{fig:fireball}.

Although we expect that energy loss due to QCD brown muck
is efficient, a significant fraction of quirks annihilate
after only a few crossings (see \Eq{PannQCD}).
The branching ratio for these annihilations into leptons
or photons are much smaller than the decays above
(see \Eqs{annleptonbr} and \eq{annphotonbr}),
but may be worth searching for.
The width of this enhancement is of order $\mq$,
and the shape is determined from the 2-particle invariant
mass distribution of the produced quirks.
This gives an additional handle on these events.

\subsection{Non-perturbative Infracolor Interactions}
We now consider non-perturbative infracolor interactions of the quirks.
There are many analogies with the non-perturbative QCD interactions
of colored quirks discussed in the previous subsection, so our discussion
will be brief and highlight the important differences.

The infracolor ``brown muck'' has a geometrical cross section for
interaction, so we also expect $\sim 1$ interaction per classical
crossing time.
As argued in Subsection~\ref{subsec:stringprod},
radiation of infracolor glueballs takes place only while
the quirk separation is less than or of order $\La^{-1}$.
The non-perturbative infracolor interactions will therefore
give rise to the emission of only $\sim 1$
infracolor gluons with total energy $\sim \La$.

One important difference with the QCD case is that the infracolor
hadrons generally do not interact after they are emitted,
and therefore their angular position is probably not ``measured''
on time scales relevant for colliders.
The cross section for an infracolor glueball with energy
$\sim \La$ to scatter \eg\ via $\ga g \to \ga g$ is of order
\beq
\si \sim \frac{1}{16\pi} \frac{\La^6}{\mq^8}
\sim 10^{-16} \si_W \left( \frac{\La}{\mbox{GeV}} \right)^4
\left( \frac{\mq}{\mbox{TeV}} \right)^{-8},
\eeq
where $\si_W \sim \La^2 / 16\pi M_W^2$ is a typical weak
cross section.
However, even if we assume that quantum coherence is maintained
between the angular wavefunction and the wavefunction of the
emitted infracolor hadrons, we still expect the probability
to find the quirks in a low partial wave after $N$ interactions
to go like $1/\sqrt{N}$,
since the quirk wavefunction is ``random-walking'' away from
low partial waves with each interaction.
We therefore expect these interactions to suppress annihilation
similarly to the QCD case.

\begin{figure}[t!]
\begin{center}
\includegraphics[]{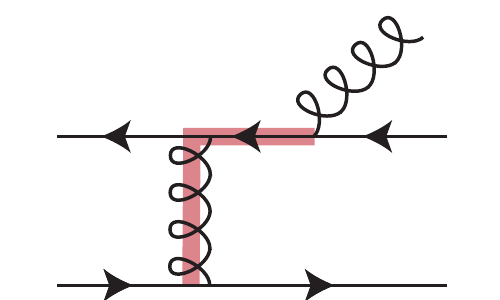}
\begin{minipage}[t]{5in}%
\caption[]{Diagram contributing to infracolor
energy loss at high momentum transfer.
Hard internal lines are shaded.}
\label{fig:ICmuckdiagram}
\end{minipage}
\end{center}
\end{figure}

Another potentially important difference from the QCD case is the 
fact that the glueball mass is 
of order the strong interaction scale,
so it is possible that glueball emission is kinematically suppressed.
For example, lattice simulations of $SU(3)$ gauge theory indicate
that the mass of the $0^{++}$ glueball is $3.6$ times heavier than
the square root of the string tension \cite{Teper}.
Although there is no parametric suppression, one should keep in mind
the possibility that there is some kinematic suppression of glueball production.
The amplitude to emit a hard infracolor gluon is shown in
Fig.~\ref{fig:ICmuckdiagram}.
The amplitude has one off-shell gluon and one off-shell heavy quirk
line, and therefore is suppressed by $1/q^3$ where $q$ is the hard
momentum transfer, so the cross section is down by $1/q^6$
at large $q$.
If this behavior sets in already
at the glueball mass, we can imagine
a suppression of order $(\frac 13)^6 \sim 10^{-3}$ in the cross
section.

We therefore consider the two extreme scenarios:
one where there is no suppression,
and one where non-perturbative infracolor interactions are effectively
absent.
In the first case, there can be significant energy loss to
infracolor gluons for sufficiently large $\La$, which gives
rise to unobservable missing energy.
In the second case, other mechanisms of energy loss (\eg\ radiation)
may be important.

\subsection{Magnetic Field}
Another effect that can be important
in preventing annihilation is the magnetic field in the detector,
of order Tesla at the Tevatron and LHC.
The quirk center of mass frame is boosted relative to the
lab frame, so there will be an electric field in this frame.
This electric field will typically have a component perpendicular
to the direction of the quirk motion, which will give rise
to a repulsive force between oppositely charged quirks.
This in turn will generate a classical separation after 
one oscillation of order
\beq
\De r \sim a T^2
\sim \frac{v_{\rm cm} B}{\mq} \left( \frac{\mq}{\La^2} \right)^2.
\eeq
where $v_{\rm cm}$ is the velocity of the center of mass
frame.
Demanding that this is larger than $\mq^{-1}$ 
(and assuming $v_{\rm cm} \sim 1$) gives
\beq
\La \lsim \mbox{MeV}
\left( \frac{B}{\mbox{Tesla}} \right)^{1/4}
\left( \frac{\mq}{\mbox{TeV}} \right)^{1/2}.
\eeq
Since magnetic fields at Tevatron and LHC colliders
are of order Tesla, we expect that this will prevent
re-annihilation for $\La \lsim \mbox{MeV}$.

This mechanism will be ineffective for special kinematical
configurations.
The induced electric field vanishes if the center of mass
of the system is along the magnetic field, that is the
beam direction.
The induced electric field does not give rise to a transverse
separation between the quirks if their motion in the center
of mass frame is along the magnetic field.
It is straightforward to check that the magnetic
field in the center of mass frame does not cause a transverse
quirk separation to leading order in the magnetic field.
Therefore, if the magnetic field is the only effect
preventing annihilation, there may be some events
in corners of kinematic phase space that annihilate.

\subsection{Electromagnetic Radiation}
The rate of electromagnetic radiation can be estimated from the
Larmor formula
\beq
\dot{E} \sim \al (\ddot{d})^2,
\eeq
where $d$ is the dipole moment of the charge distribution
(with the charge factored out).
A perfect $s$-wave has $d \equiv 0$, but even if the total
angular momentum is $\ell \sim 1$ the dipole moment will be
of order $r_{\rm max}$.
The energy radiated in a crossing time $T$ is therefore
\beq
\De E \sim \dot{E} T \sim \al \left(\frac{r_{\rm max}}{T}\right)^2
\sim \frac{\al}{T}.
\eeq
Since the typical photon energy radiated is of order 
$E_\ga \sim 1/T$, this means that there are of order $\al$
photons emitted in each classical crossing.
The number of crossings required to lose energy of order
$\mq$ to electromagnetic radiation is therefore of order
\beq
N_\ga \sim \frac{\mq}{\dot{E}} \frac{1}{T}
\sim \frac{\mq^2}{\al \La^2}
\sim 10^8 \left( \frac{\La}{\mbox{GeV}} \right)^{-2}
\left( \frac{\mq}{\mbox{TeV}} \right)^{2}.
\eeq
This is much larger than the number of crossings to
annihilate.
Brown muck interactions can prevent annihilation, but then
they will be the dominant energy loss mechanism
since $E_\ga \ll \La$.
We conclude that electromagnetic energy loss is unlikely
to be important.

\section{Mesoscopic Strings}
\label{sec:mesoscopic}
We now consider the case where the strings are too small to be
resolved in a detector (roughly $L \lsim \mbox{mm}$), but are
large compared to atomic scales ($L \gsim \mbox{\AA}$).
This corresponds to roughly
\beq
10~\mbox{keV} \lsim \La \lsim \mbox{MeV}
\eeq
for $\mq \sim \mbox{TeV}$.
In this case, the quirk-antiquirk pair will appear
as a single particle in the detector.

For mesoscopic strings, we can no longer take for granted that
matter interactions will randomize the angular momentum
and prevent the quirks from annihilating.
In other words, we need to know whether the bound state
lives long enough to appear in the detector.
The interaction region has an inner radius of
order cm with very high vacuum, and matter interactions are
not important there.
We must therefore consider other mechanisms to prevent
annihilation.

An important effect in preventing annihilation is the magnetic
field.
In the quirk center of mass frame, there will be an electric
field with a component perpendicular to the direction of quirk
motion that gives rise to a separation of classical quirk
trajectories after one oscillation of order
\beq
\De r \sim a T^2
\sim \frac{v_{\rm cm} B}{\mq} \left( \frac{\mq}{\La^2} \right)^2.
\eeq
where $v_{\rm cm}$ is the velocity of the center of mass
frame.
Demanding that this is larger than $\mq^{-1}$ 
(and assuming $v_{\rm cm} \sim 1$) gives
\beq
\La \lsim \mbox{MeV}
\left( \frac{B}{\mbox{Tesla}} \right)^{1/4}
\left( \frac{\mq}{\mbox{TeV}} \right)^{1/2}.
\eeq
Since magnetic fields at Tevatron and LHC colliders
are of order Tesla, we expect that this will prevent
re-annihilation for $\La \lsim \mbox{MeV}$.

While the quirk pair is inside the beam pipe,
the only efficient mechanism for energy loss
and change of angular momentum is the brown-muck interactions
discussed in the previous section.
For colored quirks, these lead to a decay length
(see \Eq{brownmuckdisplace})
\beq
c\tau \sim \mbox{cm}
\left( \frac{\La}{100~\mbox{keV}} \right)^{-2}
\left( \frac{\mq}{\mbox{TeV}} \right)^2
\eeq
while for uncolored quirks
\beq
c\tau \sim 10~\mbox{cm}
\left( \frac{\La}{\mbox{MeV}} \right)^{-3}
\left( \frac{\mq}{\mbox{TeV}} \right)^2.
\eeq
We see that these decays can allow the quirk bound state
to survive for distances of order cm.
As discussed in the previous section,
the efficiency of this mechanism of energy loss is uncertain,
particularly for the infracolor energy loss.
The decay lengths may therefore be significantly longer
than these estimates.

Once the bound state reaches the beam pipe, matter interactions
are efficient at randomizing the angular momentum and
preventing annihilation.
For example, a single collision with an electron
transfers momentum of order $m_e$, which changes
the angular momentum by 
\beq
\De \ell \sim m_e L \sim m_e \frac{\mq}{\La^2}
\sim 10^3 \left( \frac{\La}{\mbox{MeV}} \right)^{-2}
\left( \frac{\mq}{\mbox{TeV}} \right).
\eeq

For the remainder of this section we will assume that the
bound state appears
as a stable particle in the detector.
In order to see the bound state, it must be produced in
association with a hard jet or photon so that the bound
state is off the beam axis.
If the bound state has a net electromagnetic charge, it will leave
a track in the detector.
The signal is then a single heavy stable particle recoiling
against a hard jet or photon.

The most interesting aspect of these events is the fact that the
mass of the bound state is the invariant mass of the quirk-antiquirk
pair.
This has a broad distribution,
so the mass of the bound state differs by order 1 event by event.
The mass of a heavy stable charged
particle can be measured event by event
by a combination of its bending in a magnetic field
and time of flight.
This has been studied at LHC \cite{StableLHC},
with the conclusion that the mass can be determined at the
few percent level for events with $0.6 \lsim \be \lsim 0.8$.
Observation of stable particles with the mass spectrum
given by the 2-particle invariant mass spectrum would be
essentially a direct observation of strings.

\section{Microscopic Strings}
\label{sec:micro}
We now consider the signals for microscopic strings, roughly
$L \lsim \mbox{\AA}$, corresponding roughly to
\beq
\mbox{MeV} \lsim \La \lsim \mq / \mbox{few}.
\eeq
As we have seen above, in this regime interactions with
matter and the magnetic field of the detector do not
prevent the quirks from annihilating.
These signals have been largely discussed in Section~\ref{sec:reann},
so our discussion here is mainly a summary of this discussion.

\subsection{Colored Quirks}

We begin with colored quirks, which are the ones most
copiously produced at a hadron collider.
Most of the quirks produced above threshold
will undergo hard annihilation without significant energy
loss.
Colored quirks will annihilate dominantly into jets, but may
have branching fractions into leptons or photons at the percent
level (see Subsection~\ref{subsec:reannrate}).
These events will have a broad distribution essentially given
by the perturbative 2-particle invariant mass spectrum of the
quirks.

A significant fraction (a few percent) of colored
quirk pairs will lose most of their kinetic energy
energy because of interactions of the
non-perturbative QCD and/or infracolor interactions.
The condition for non-perturbative QCD interactions to
dominate is \naively $\LQ > \La$, but there is significant
uncertaintly in the efficiency of the non-perturbative
infracolor energy loss.
If the QCD interactions dominate, an energy of order
$2\mq$ will be radiated as light QCD hadrons (mainly pions)
each with energy of order GeV: a hadronic fireball.

The total invariant mass distribution of the quirk
decay products is therefore a broad distribution
approximating the 2-particle invariant mass 
distribution with a narrow peak superimposed.
This distribution can be found in jets,
but also (with reduced rate) in lepton or
photon pairs.
Detailed study of this signal would be very
interesting.

\subsection{Uncolored Quirks}
Uncolored electromagnetically charged quirks will annihilate
dominantly into infracolor glueballs
for the range of $\La$ of interest.
Infracolor glueballs are unobservable unless $\La \gsim 50\GeV$
(see \Eq{icglueballdecay}),
in which case they decay inside the detector.
However, there will be branching ratio typically
of order $10\%$ for annihilation into visible states
(see \Eqs{PannQED} and \eq{Panninv}).
We can have annihilation to photon pairs, or through a
virtual photon, $Z$, or $W$.
The final state will therefore include a significant fraction of
photon and lepton pairs, which are readily observable.

Non-perturbative infracolor interactions
will tend to bring the quirks to rest before
they annihilate.
If these are fully efficient, they will radiate
of order one infracolor glueball with energy of order $\La$
and total angular momentum of order 1
once per classical crossing time.
However, glueball masses are themselves of order $\La$,
so it is possible that there is a kinematic suppression
of this process.
Given our lack of understanding of this non-perturbative
dynamics, it makes sense to consider both the case where
these interactions are efficient and inefficient.

Hard annihilation of excited uncolored quirks requires of order
$10^2$ crossings (see \Eq{PannQED}), giving the non-perturbative
infracolor interactions plenty of time to randomize the angular
momentum.
If these interactions are efficient, we therefore expect the
majority of these annihilations to take place with the quirks
at rest, leading to a very narrow resonance.
If these interactions are inefficient, the resonance
will be broad with a narrow peak superimposed from the
quirks that are produced near threshold.

\section{Conclusions}
\label{sec:conclude}
We have seen that massive particles charged under
an unbroken non-abelian gauge group give rise to
spectacular phenomenology at colliders.
These signals are sufficiently exotic that they will almost
certainly be missed unless they are searched for.
Given the simple nature of these models, it is worthwhile
to put some effort in this direction.
The next step will be to produce event generators for
this exotic physics that can be used to develop concrete search
strategies.
Cosmological aspects of these models will also be addressed
in a future publication.

\newpage
\section*{Acknowledgements}
This paper had a long gestation during a
eventful and sometimes difficult time in the life
of MAL.
He would like to dedicate his work on this paper
to the memory of Alexander M.\ Luty.
It is a pleasure to thank the following colleagues for
useful discussions and (in most cases) encouragement:
Z.\ Chacko,
H.-C.\ Cheng,
A.\ De Roeck,
R.\ Harnik,
H.-S.\ Goh,
S.\ Nasri,
Y.\ Nomura,
M.\ Peskin,
M.\ Strassler,
and
J.\ Terning.
We thank J. Evans for generating the quirk production plots.
This work was partly performed at the Aspen Center for Physics
and the Kavli Institue for Theoretical Physics.
This work was partly supported by NSF grant PHY-0652363
and by the Maryland Center for String and Particle Theory.


\end{document}